\documentclass[twoside,twocolumn,9pt]{article}
\usepackage{extsizes}
\usepackage[super,sort&compress,comma]{natbib} 
\usepackage[version=3]{mhchem}
\usepackage[left=1.5cm, right=1.5cm]{geometry}
\usepackage{balance}
\usepackage{times,mathptmx}
\usepackage{sectsty}
\usepackage{graphicx} 
\usepackage{lastpage}
\usepackage[format=plain,justification=justified,singlelinecheck=false,font={stretch=1.125,small,sf},labelfont=bf,labelsep=space]{caption}
\usepackage{float}
\usepackage{fancyhdr}
\usepackage{fnpos}
\usepackage[english]{babel}
\addto{\captionsenglish}{%
  
}
\usepackage{array}
\usepackage{droidsans}
\usepackage{charter}
\usepackage[T1]{fontenc}
\usepackage[usenames,dvipsnames]{xcolor}
\usepackage{setspace}
\usepackage[compact]{titlesec}
\usepackage{hyperref}
\usepackage{epstopdf}

\definecolor{cream}{RGB}{222,217,201}

\begin{document}

\pagestyle{fancy}
\thispagestyle{plain}
\fancypagestyle{plain}{

\fancyhead[C]{\includegraphics[width=18.5cm]{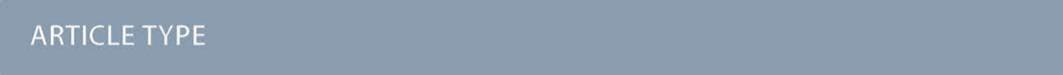}}
\fancyhead[L]{\hspace{0cm}\vspace{1.5cm}\includegraphics[height=30pt]{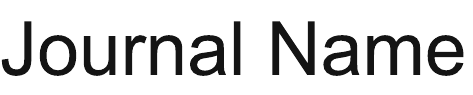}}
\fancyhead[R]{\hspace{0cm}\vspace{1.7cm}\includegraphics[height=55pt]{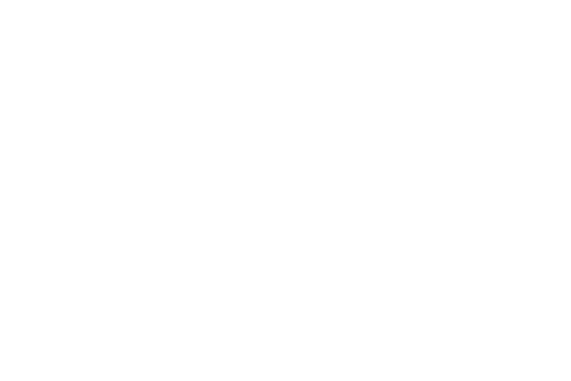}}
\renewcommand{\headrulewidth}{0pt}
}

\makeFNbottom
\makeatletter
\renewcommand\LARGE{\@setfontsize\LARGE{15pt}{17}}
\renewcommand\Large{\@setfontsize\Large{12pt}{14}}
\renewcommand\large{\@setfontsize\large{10pt}{12}}
\renewcommand\footnotesize{\@setfontsize\footnotesize{7pt}{10}}
\makeatother

\renewcommand{\thefootnote}{\fnsymbol{footnote}}
\renewcommand\footnoterule{\vspace*{1pt}%
\color{cream}\hrule width 3.5in height 0.4pt \color{black}\vspace*{5pt}} 
\setcounter{secnumdepth}{5}

\makeatletter 
\renewcommand\@biblabel[1]{#1}            
\renewcommand\@makefntext[1]%
{\noindent\makebox[0pt][r]{\@thefnmark\,}#1}
\makeatother 
\renewcommand{\figurename}{\small{Fig.}~}
\sectionfont{\sffamily\Large}
\subsectionfont{\normalsize}
\subsubsectionfont{\bf}
\setstretch{1.125} 
\setlength{\skip\footins}{0.8cm}
\setlength{\footnotesep}{0.25cm}
\setlength{\jot}{10pt}
\titlespacing*{\section}{0pt}{4pt}{4pt}
\titlespacing*{\subsection}{0pt}{15pt}{1pt}

\fancyfoot{}
\fancyfoot[LO,RE]{\vspace{-7.1pt}\includegraphics[height=9pt]{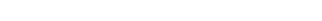}}
\fancyfoot[CO]{\vspace{-7.1pt}\hspace{13.2cm}\includegraphics{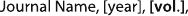}}
\fancyfoot[CE]{\vspace{-7.2pt}\hspace{-14.2cm}\includegraphics{head_foot/RF}}
\fancyfoot[RO]{\footnotesize{\sffamily{1--\pageref{LastPage} ~\textbar  \hspace{2pt}\thepage}}}
\fancyfoot[LE]{\footnotesize{\sffamily{\thepage~\textbar\hspace{3.45cm} 1--\pageref{LastPage}}}}
\fancyhead{}
\renewcommand{\headrulewidth}{0pt} 
\renewcommand{\footrulewidth}{0pt}
\setlength{\arrayrulewidth}{1pt}
\setlength{\columnsep}{6.5mm}
\setlength\bibsep{1pt}

\makeatletter 
\newlength{\figrulesep} 
\setlength{\figrulesep}{0.5\textfloatsep} 

\newcommand{\topfigrule}{\vspace*{-1pt}%
\noindent{\color{cream}\rule[-\figrulesep]{\columnwidth}{1.5pt}} }

\newcommand{\botfigrule}{\vspace*{-2pt}%
\noindent{\color{cream}\rule[\figrulesep]{\columnwidth}{1.5pt}} }

\newcommand{\dblfigrule}{\vspace*{-1pt}%
\noindent{\color{cream}\rule[-\figrulesep]{\textwidth}{1.5pt}} }

\makeatother

\twocolumn[
  \begin{@twocolumnfalse}
\vspace{3cm}
\sffamily
\begin{tabular}{m{4.5cm} p{13.5cm} }

\includegraphics{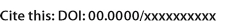} & \noindent\LARGE{\textbf{On the fundamentals of quantum rate theory and the long-range electron transport in respiratory chains$^\dag$}} \\
\vspace{0.3cm} & \vspace{0.3cm} \\

& \noindent\large{Paulo Roberto Bueno,$^{\ast}$\textit{$^{a}$}} \\

\includegraphics{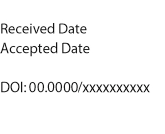} & \noindent\normalsize{It has been shown that both the electron-transfer rate constant of an electrochemical reaction and the conductance quantum are correlated with the concept of quantum capacitance. This simple association between the two separate concepts has an entirely quantum rate basis that encompasses the electron-transfer rate theory as originally proposed by Rudolph A. Marcus~\citep{Bueno-2020-ET} whether statistical mechanics is properly taken into account. Presently, it is conducted a concise review of the quantum mechanical rate theory principles focused on its quantum electrodynamics character to demonstrate that it can reconcile the conflicting views established on attempting to use the super-exchange (supported on electron transfer) or `metallic-like' (supported on conductance quantum) mechanisms separately to explain the highly efficient long-range electron transport observed in the respiratory processes of living cells. The unresolved issues related to long-range electron transport are clarified in light of the quantum rate theory with a discussion focused on \textit{Geobacter sulfurreducens} films as a reference standard of the respiration chain. Theoretical analyses supported by experimental data suggest that the efficiency of respiration within a long-range electron transport path is intrinsically a quantum mechanical event that follows relativistic quantum electrodynamics principles as addressed by quantum rate theory.}

\end{tabular}

\end{@twocolumnfalse} \vspace{0.6cm}

]

\renewcommand*\rmdefault{bch}\normalfont\upshape
\rmfamily
\section*{}
\vspace{-1cm}


\footnotetext{\textit{$^{a}$~Institute of Chemistry, Sao Paulo State University, Araraquara, Sao Paulo, Brazil. E-mail: paulo-roberto.bueno@unesp.br.}}


\section{Introduction}\label{sec:introduction}

\begin{figure}[!t]
\centering
\includegraphics[scale=0.45] {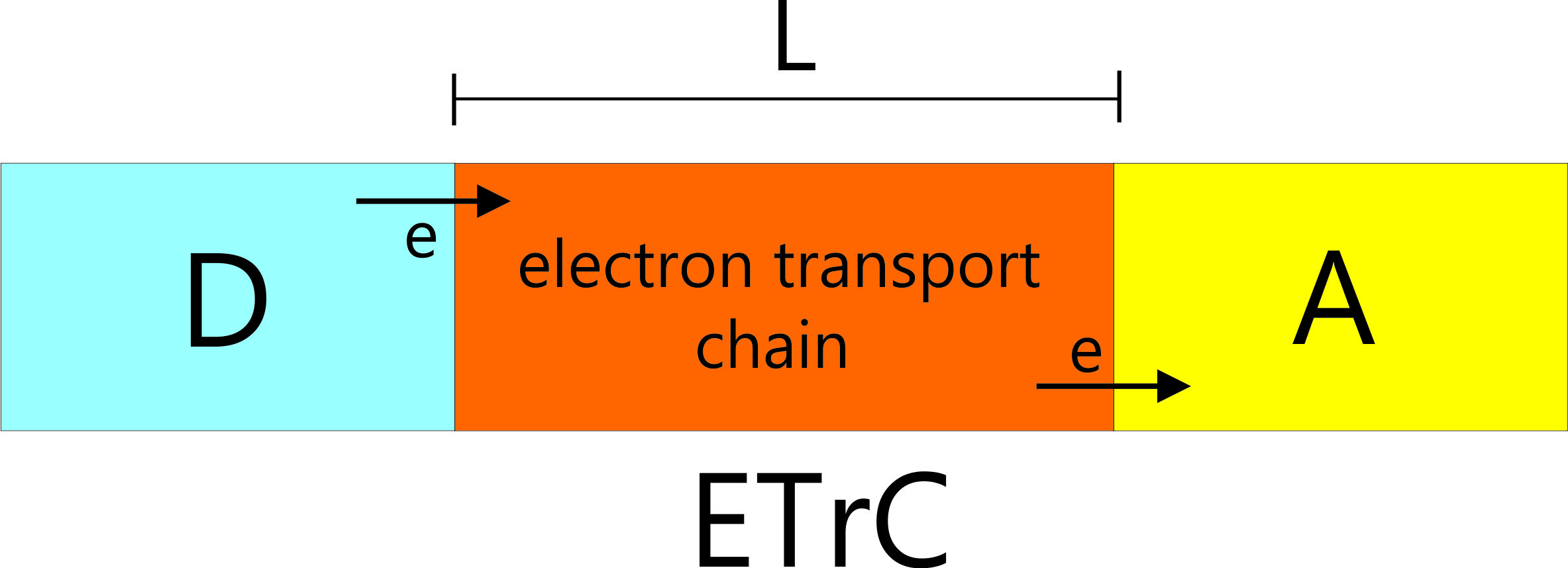}
\caption{Typical set-up for measuring the transport of electrons in biological systems. The electric current that flows from $D$ to $A$ states throughout an electron transport chain (ETrC) which generally, in biological systems, contains redox proteins or metal-containing cofactors. The length $L$ of the ETrC, indicated in this figure, is within a micro to millimetre scale.}
\label{fig:ETrC-scheme}
\end{figure}

The energy conservation of living cells\cite{Kushnareva-2002, Brennan-2010, Engel-2007}, and consequently of all living organisms, requires the transport of electrons\cite{Engel-2007} within respiratory chains \cite{Logan-2009} and photosynthetic processes \cite{Lewis-2006, Brennan-2010}. It is an experimental fact that the transport of electrons in biological systems occurs through a long-range length scale pathway~\citep{Kaila-2018}, which is generally denoted as electron transport chain (ETrC)~\citep{Ramsay-2019}. This long pathway of length $L$ separates donor \textit{D} and acceptor \textit{A} energy state levels, as illustrated in Figure~\ref{fig:ETrC-scheme}, and can be longer than a few micrometres and reaches, in some exceptional circumstances, remarkable sizes of millimetres~\citep{Yang-NC-2021}. 

\begin{figure}[!t]
\centering
\includegraphics[scale=0.55] {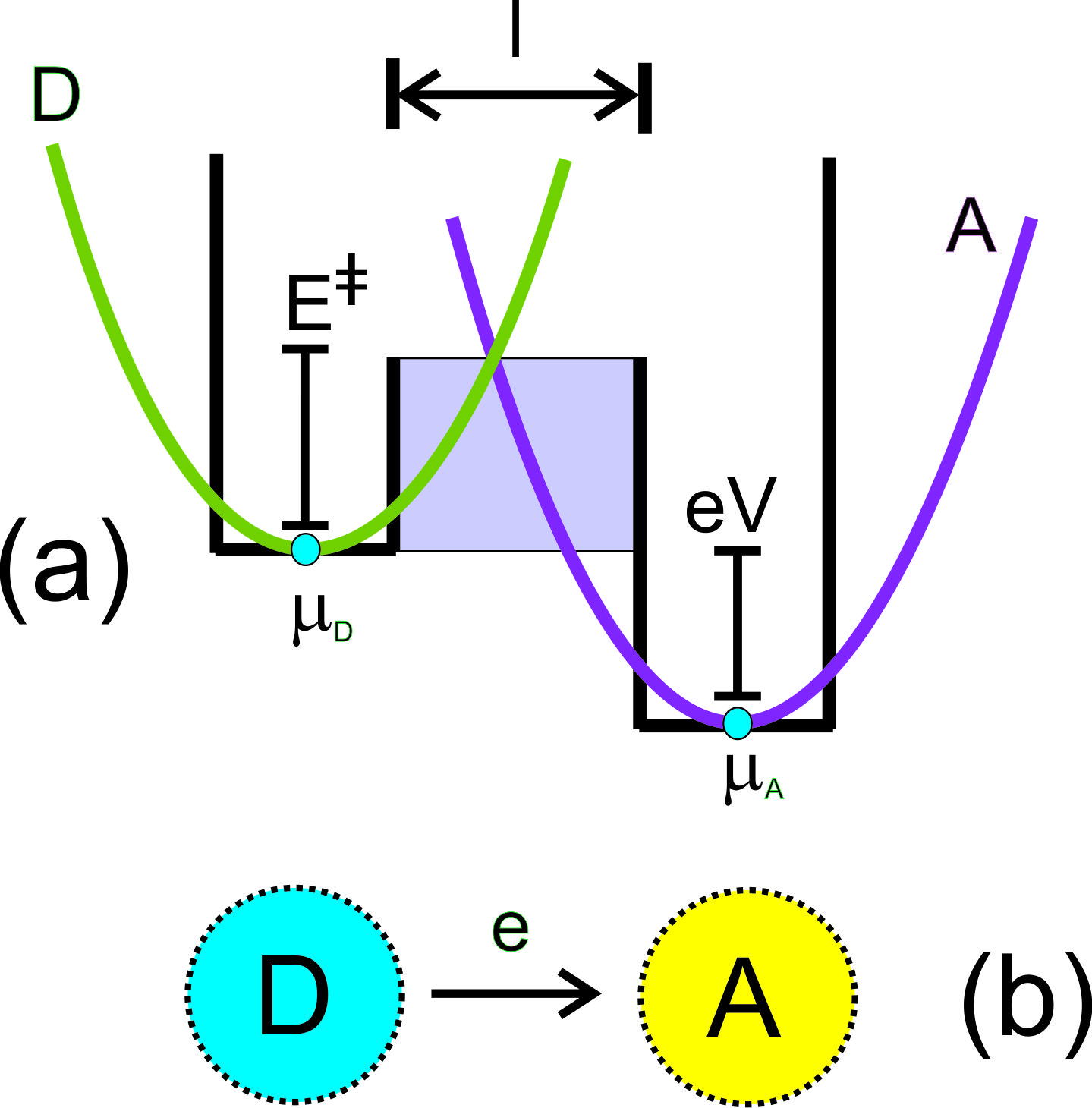}
\caption{(a) $D$ and $A$ states comprising quantum channel with short path length $l$ (no longer than 3 nm) for ET through a potential energy of the barrier ($eV$) that defines a tunneling electron coupling between $D$ and $A$ such that $\kappa = \exp \left( {-\beta l} \right)$, where $\beta$ is a constant that defines the properties of the barrier; and $\mu_{D}$ and $\mu_{A}$ are the chemical potentials states $D$ and $A$, defining a chemical potential difference that complies with $\Delta \mu = -eV = e^2/C_q$~\citep{Sanchez-2022-1, Bueno-2023-3}. For an adiabatic or ballistic quantum conductance with a perfect transmittance $\kappa \sim 1$, the conductance through the barrier is theoretically stated as $G_0 = g_s e^2/h \sim$ 77.5 $\mu$S~\citep{Sanchez-2022-1, Bueno-2023-3}. (b) $D$ (oxidation) and $A$ (reduction) states of the redox reaction Ox + e $\rightharpoonup$ Red has been demonstrated to possess electrodynamics that follows relativistic quantum mechanics~\citep{Sanchez-2022-2, Bueno-2023-3}.}
\label{fig:general-sch-QRT}
\end{figure}

\begin{figure}[!t]
\centering
\includegraphics[scale=0.45] {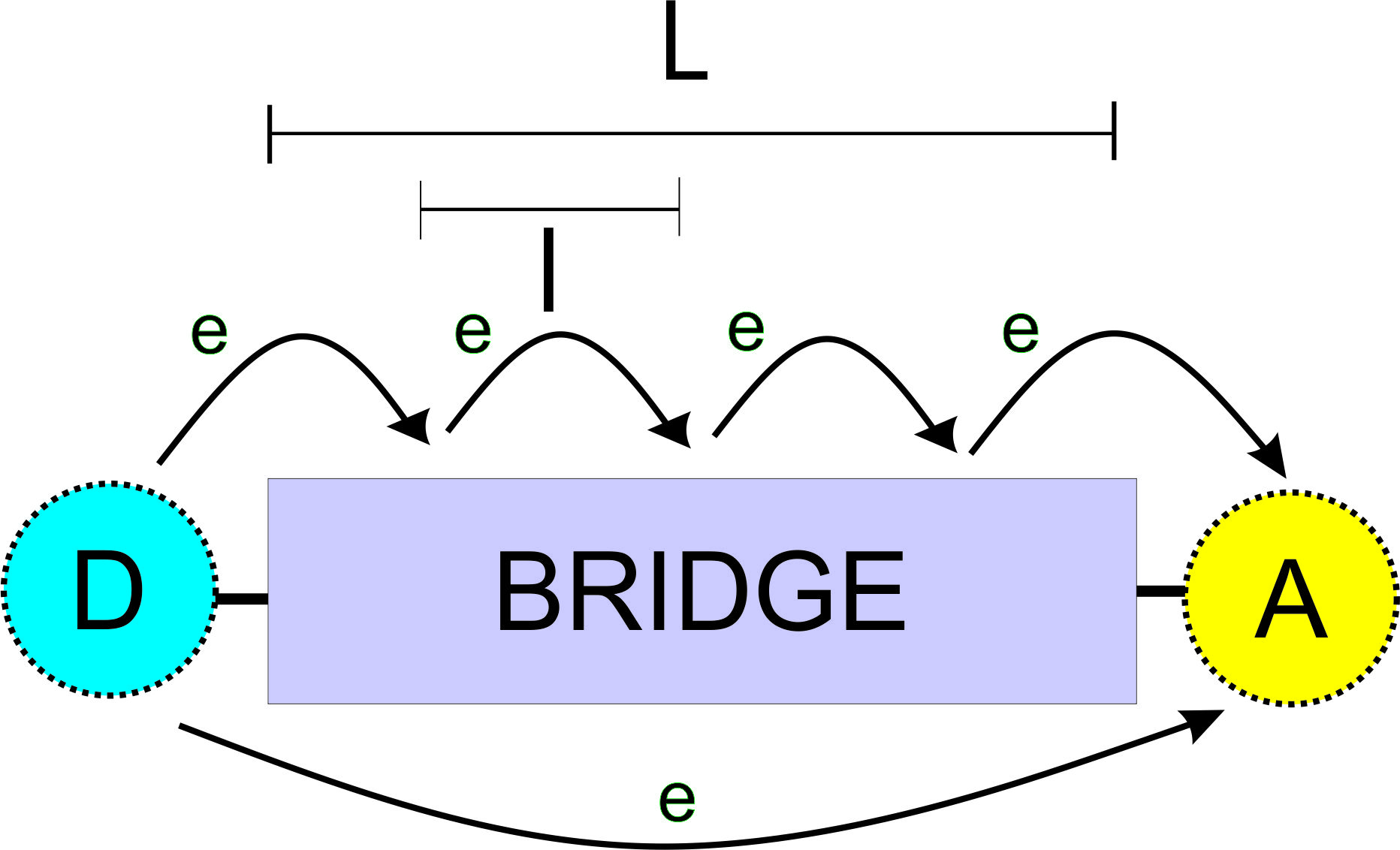}
\caption{Illustration of a $D$-bridge-$A$ structure comprising $D$ and $A$ chemically coupled to a bridge molecular structure. The transport of charge (electrons in the present case) through the molecular bridge structure can be conducted by a direct charge transport through a single longer path $L$ (referred to as super-exchange mechanism) (bottom of the figure) or by a sequential ET mechanism (top of the figure) through shorter paths. Presently, the direct or sequential ET via molecular bridge or tunnelling will be referred indistinctly as super-exchange mechanism.}
\label{fig:D-B-A-structure}
\end{figure}

The differences between the scheme of Figure~\ref{fig:ETrC-scheme} and the electron transfer\footnote{For the sake of simplicity, electron transfer in the present text will be used as a synonym of charge transfer, hence phenomenologically incorporating the transfer of electron and hole as carriers through potential barriers or molecular bridge structures equivalently.} (ET) between $D$ and $A$ in redox homogeneous reactions (as schematically depicted in Figure~\ref{fig:general-sch-QRT}) or as intermediated by chemical bridges in intra-molecular ET (as depicted in Figure~\ref{fig:D-B-A-structure}) is not only owing to the physical properties of the potential barrier (in the case of redox homogeneous ET) or the chemical nature of the molecular bridge (in the case of intra-molecular ET), which ultimately depends, in all cases, on the length separating $D$ and $A$ states. To better understand the correspondence between ETrC (Figure~\ref{fig:ETrC-scheme}) and ET mechanisms (Figures~\ref{fig:general-sch-QRT} and ~\ref{fig:D-B-A-structure}), it requires differentiate between the concepts of ET and electron transport, which frequently has been used indistinctly or wrongly as synonymous, although there are clear conceptual differences between these two phenomena~\citep{Bueno-2020-ET}.

\begin{figure}[!t]
\centering
\includegraphics[scale=0.55] {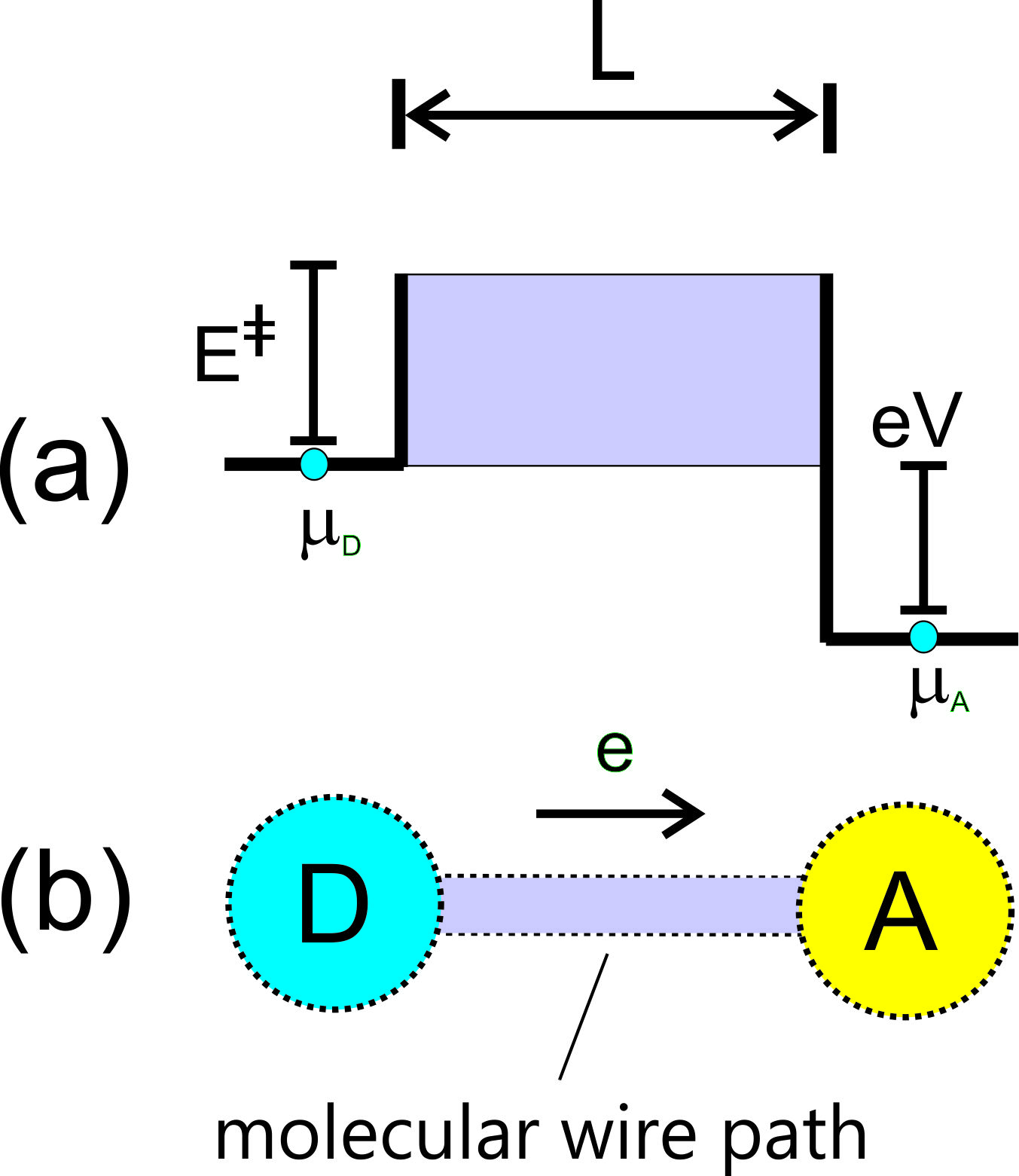}
\caption{(a) Within the quantum rate theory, there are few differences between the scheme of ET as depicted in Figure~\ref{fig:general-sch-QRT}, as the basement of a sequential super-exchange tunneling ET mechanism, and the long-range transport of electrons depicted here to occurs through longer path $L$ with the assistance of a molecular wire path structure. Nonetheless, within the description of the quantum rate model, the differences are only associated with the existence of activation energy in the ET scheme (see Figure~\ref{fig:general-sch-QRT}\textit{a}) whereas this energy term is absent in the (b) ballistic (coherent) or incoherent hopping long-range transport of electrons through a nanowire of length $L$. These differences, in the quantum rate model, are quantitatively predicted simply through the nature of the statistical distribution of energy $E = e^2/C_q$ and transmission coefficient $\kappa$ term in the constitutive equations of the model.}
\label{fig:QRT-ballistic-Tr}
\end{figure}

Particularly, ET was historically formulated as a rate~\citep{Marcus-1993} that determines the velocity of redox homogeneous reaction~\cite{Bueno-2023-3} or the frequency of electron exchanged between $D$ and $A$ states in conjugated organic molecules~\cite{Bueno-2023-ICT} comprising electron-donating and electron-withdrawing groups most commonly separated by a molecular bridge comprised (but not limited to) of $\pi$-bridge chemical structures whereas electron transport requires knowing the mechanism of electron conductance operating in the molecular bridge (which is referred to molecular wire~\citep{Imry-Landauer-1999} whenever the mechanism of electron transport in the bridge is ballistic, see Figure~\ref{fig:QRT-ballistic-Tr}\textit{b}). Both ET and electron transport concepts are usually applied indistinctly to describe the transport of electrons owing to they are related by a particular setting of another key physical concept referred to as electrochemical capacitance~\cite{Bueno-2020-ET, Bueno-2023-3}, as will be demonstrated in section~\ref{sec:QRTheory}.

Note that to determine the ET rate, it is required to know the energy level differences between $D$ and $A$ states and the electronic coupling established by a dielectric medium (in the case of homogeneous ET reactions) or molecular bridge structure (in the case of Figures~\ref{fig:D-B-A-structure} and~\ref{fig:QRT-ballistic-Tr}) separating these states. The separation path between $D$ and $A$, in a homogeneous redox reaction, occurs via solvent/electrolyte dielectric-type environment is short (where the ET mechanism is governed by electron tunnelling, as noted in Figure~\ref{fig:general-sch-QRT}) whereas in molecular bridge structures (including molecular wires) it can be short or long, established, for instance, via molecular-bridge structures and operates via intra-molecular ET mechanism, which is governed by super-exchange, or sequential electron hopping, as noted in Figure~\ref{fig:D-B-A-structure}. 

In biological processes (e.g., reduction in RNA to DNA, photosynthesis, and aerobic and anaerobic respiration), the path or the environment properties (that necessarily requires the presence of aqueous -- dielectric solvent -- electrolyte) established for the electron transport have not been a limitation for the efficient transmission of electrons between initial $D$ and final $A$ energy levels despite of the long-range length of the ETrC. The problem with the understanding of the efficiency of electron transport in a long-range pathway, as will be demonstrated here, is related to a comprehensive knowledge of the correlation between ET rate, quantum conductance, and quantum capacitance concepts within an electrolyte environment~\citep{Bueno-2020-ET}.

The efficiency of charge transport in the long-range length of the ETrC has been attributed to the super-exchange or sequential (charge ET hopping) conductive mechanisms which are when an electron (or hole\footnote{The physical differences between electron and hole as charge carriers under a bias electric or chemical potential differences are owing to that electrons have an elementary negative charge magnitude and holes have positive within different effective masses for the transport of their charge. The differences in the effective masses of these types of charge carriers are associated with the mobility of each carrier in the chemical environment associated with the electronic structure of the compounds, with free electrons in a conduction band environment, for instance, occupying a higher energy level (thus possessing lower effective mass and higher mobility) whereas holes, in the valence band, occupying lower (hence possessing higher effective mass and lower mobility).}) is transferred from $D$ to $A$ state with the assistance of an intermediate group acting as a molecular bridge between $D$ and $A$ states. As illustrated in Figure~\ref{fig:D-B-A-structure}, the super-exchange is the direct, long-distance\footnote{The term long-distance refers here to a distance in between 3 to 20 nm whereas long-range pathway (that comprises the definition of ETrC in biological structures, as illustrated in Figure~\ref{fig:ETrC-scheme}) describes the transport of charge in the micro or millimetre scale.} electron transport enhanced by the constructive interference of $D$ and $A$ wave functions which in turn occurs by the superposition of $D$ and $A$ molecular orbitals. In the sequential ET mechanism, as a form of long-distance charge transport, the charge temporarily resides on the midway, molecular bridge group (see Figure~\ref{fig:D-B-A-structure}), while in the super-exchange, this intermediate state only participates by providing a quantum channel\footnote{The term quantum channel will be defined later during the introduction of the quantum rate theory path for the transport of the charge.} for the transmittance of the electron.

Although advances in theory and experiment during the last sixty years have been able to elucidate the electron transport in chemistry and biology structures~\citep{Matoba-2006, Murphy-2009}, many fundamental problems remain unsolved regarding the long-range pathway issue. It has been accepted that the super-exchange is dominant, but not the exclusive $D$-$A$ coupling mechanism for long-range ET. The acceptance of the super-exchange mechanism occurs even though mediating states and energy gaps are rarely identified and are not well-correlated with spectroscopic and thermodynamic properties of ETrC bridging medium and hence are unable to resolve between coherent electron tunnelling and incoherent hopping mechanisms~\citep{Winkler-2014, Gray-2005}. 

It is difficult to distinguish and, consequently, to experimentally resolve between coherent \textit{versus} incoherent electron transport. Essentially, coherent electron transport consists of the existence of a phase coherence owing to a constant phase difference existing between, for instance, a potential bias difference and an associated measurable electric current or waves of equivalent frequency. For instance, electrons can behave as waves in the case of a ballistic mechanism of transport~\citep{Imry-Landauer-1999}, which is defined in terms of the medium in which electrons are being transported. The ballistic transport is thus dependent on the electrical resistance/conductance of the medium which necessarily requires a minimal or negligible electron scattering caused by the atoms, molecules, or impurities in the medium itself. By definition, the Landauer resistance/conductance~\citep{Landauer-1957} (to be better introduced further here as a key component of the quantum rate theory) intrinsically incorporates a ballistic electron transport physical phenomenology~\citep{Imry-Landauer-1999}. This ballistic electron transport has been experimentally demonstrated~\citep{Bueno-2023-3} to exist superimposed to the ET tunneling mechanism in redox reaction occurring in molecular films~\citep{Bueno-2023-3} and demonstrated to operate independent of a short or long distance provided that the energy states associated with the quantum/electrochemical capacitance are energetically aligned~\cite{Sanchez-2022-2}.

A current challenge paradigm, concerning ETrC in biology, is the empirical studies involving the \textit{Geobacter sulfurreducens} respiration~\citep{Bond-2003, Strycharz-Glaven-2011, Strycharz-Glaven-2012}. This bacterial system possesses nanowires molecular structures (which is here used as synonymous with the molecular bridge in which the electron transport mechanism is essentially ballistic) where the transmission of electrons is performed by extracellular Fe$^{+3}$ oxides that are contacted via micrometer-long hairlike appendages known as pili~\citep{Shi-2016, Reguera-2005}. These pili structures are coated with multiheme \textit{c}-type cytochromes that have been suggested to serve as hopping intermediates in micrometer-distance electron-transport processes. However, an alternative viewpoint has been suggested that interprets the pilus acting as `metal-like'\footnote{Metallic-like term for describing molecular wire ballistic conductance is unfortunate. I am using it here to be consistent with the literature's debate, but I want to reinforce that ballistic conductance is physically quite different from that of metallic. Metallic conductance has a drift velocity associated with it that is orders of magnitude lower ($\sim 10^{-3}$ m s$^{-1}$) than the Fermi velocity ($\sim 10^{6}$ m s$^{-1}$) of electrons performing the ballistic translational movement in one-dimensional molecular wire structures.} conductive structures~\citep{Malvankar-2015, Malvankar-2011, Lovely-2011} in the absence of the cytochromes. In this intense debate, some authors propose a super-exchange tunneling interpretation that imposes severe constraints on long-range ET process interpretation of ballistic conductance of the pili structures~\citep{Strycharz-Glaven-2011, Malvankar-2012, Strycharz-Glaven-2012}.  However, in support of the `metal-like' conductance mechanism, more recently, it was demonstrated that pili structures are not the nanowires filament structures responsible for connecting $D$ and $A$ of the globally important redox phenomena, but surprisingly the \textit{Geobacter sulfurreducens} nanowires are assembled by micrometer-long polymerization of the multiheme c-type cytochrome, particularly the hexaheme cytochrome, referred as OmcS, with heme structures packed within $\sim$ 3 to 6 angstrons of each other\citep{Wang-2019}.

In the present review, it will be discussed and proposed an alternative interpretation sustained by a quantum mechanical first-principle rate theory~\citep{Bueno-2023-3, Bueno+Davis-2020, Bueno-book-2018}. However, before introducing what the quantum mechanical rate theory is about, it will be conducted, in the next section, a clarification on the interpretation of short- and long-range electron transport within classical ET theory, which is the basement for our current understanding of the electrodynamics of redox reactions.

\subsection{Short- and Long-range Transport of Electrons} \label{sec:short-long-etransport}

As noted in section~\ref{sec:introduction}, there are support for the transport of electrons in respiratory chains to occur through redox reaction mechanism~\cite{Li-1997, Matoba-2006, Murphy-2009, Reguera-2005}; for example, it would involve cytochrome redox states supported on the ET theory introduced by Marcus~\cite{Marcus-1985}. Marcus's theory accounts for homogeneous ET setting and has undoubtedly been crucial for understanding the bioenergetics of photosynthesis and respiration~\cite{Marcus-1985, Gust-2001}.

In the scheme of electron transport, as a result of sequential ET processes through adjacent redox sites, the distance between $D$ and $A$ states needs to be of short-range for ET to proceed (see Figure~\ref{fig:general-sch-QRT}\textit{a}), as a requirement of the electron coupling between redox sites that occurs following electron tunneling mechanisms. This scheme of the long-range electron transport through sequential short-range ET processes is denoted as a super-exchange mechanism, no matter if it operates by tunneling (Figure~\ref{fig:general-sch-QRT}) or intra-molecular bridge structures (Figures~\ref{fig:D-B-A-structure} and~\ref{fig:QRT-ballistic-Tr}). Hence, presently, note that the direct or sequential ET mechanisms will, from now on, be indistinctly referred to as super-exchange mechanism and the electron transport (coherent or incoherent), generally referred to in the literature as `metallic-like' mode of transmitting the electron will be referred here simply as electron transport, with both concepts contained and be unified by the quantum rate theory~\citep{Bueno-2020-ET, Bueno-book-2018}. Within a super-exchange mechanism the rate of the ET process $k$ is crucial and is typically modeled by the classical ET transfer rate theory, as was originally derived by Marcus as~\cite{Marcus-1964}

\begin{equation}
 \label{eq:Arrhenius-Marcus}
	k = \kappa \frac{k_BT}{h} \exp \left( \frac {E^{\ddagger}} {k_{B}T} \right),
\end{equation}

\noindent where $\kappa$, $k_{B}$, $T$, $h$, and $E^{\ddagger} = (\lambda_0 + E)^2 / 4 \lambda_0$ denote the transmission coefficient, Boltzmann constant, absolute temperature, Plank constant, and activation free energy of the ET process, respectively. In the definition of $E^{\ddagger}$, as proposed by Marcus\cite{Marcus-1964}, $E$ (represented as $eV$ in Figure~\ref{fig:general-sch-QRT}) denote the free energy of the redox reaction and $\lambda_0$ the energy of the reorganisation of the solvent\cite{Marcus-1993}, respectively.

In the classical ET theoretical approach stated in Eq.~\ref{eq:Arrhenius-Marcus}, the electron transmission coefficient term $\kappa$ ensures that the electron transport mechanism is governed by short-range (few angstroms to a maximum of 2-3 nm) electron tunneling transport between $D$ and $A$ states. Owing to this short-range individual ET scheme, the long-range transport of the electrons is inferred to be conducted by multiple sequential ET coupled processes, where $\kappa$ is the parameter that models the adiabatic character of the electron transport. The inference of multiple ET sequential steps is a critical presumption that supports the super-exchange mechanism. Pieces of evidence for this multiple ET sequential super-exchange electron transport scheme have been experimentally demonstrated~\citep{Reguera-2005}, where it is shown that the electron coupling, in ET reactions, occurs via electron tunneling modeled as $\kappa \sim \exp (-\beta l)$\footnote{Presently, it will be used $l$ to denote short-range tunneling coupled electron transport whereas $L$ will be used to denote the long-range path associated with ETrCs.}. The $\beta$ term in $\kappa \sim \exp (-\beta l)$ is a constant that defines the properties of the potential barrier for the electron to tunnel and $l$ is the short-range length of the individual sequential barriers.

In summary, the problem in explaining the long-range pathway for electron transport requires resolving the fundamental contradiction between super-exchange (supported in both adiabatic~\citep{Bueno-2023-ICT} or non-adiabatic electron transfer~\citep{Bueno-2023-3}) and coherent (supported in quantum ballistic conductance)~\citep{Eshel-2020} or incoherent (supported in sequential hopping)~\citep{Dahl-2022} conductance mechanisms. In the next section, it will be discussed the fundamental aspects that support each of these mechanisms within a respiration viewpoint by taking \textit{Geobacter sulfurreducens} films as a reference of ETrC.

\subsection{Super-Exchange \textit{versus} Electron Transport} \label{sec:ETr-mechanisms}

As shortly introduced in section~\ref{sec:short-long-etransport}, in biological structures such as respiratory chain of biological \textit{Geobacter sulfurreducens} films, in which \textit{Geobacter sulfurreducens} is an anaerobic species of bacteria that has been used as a typical biological frame of reference to study the respiration processes of living organisms~\cite{Reguera-2005, Bond-2003}, the electrons can be transported over distances of 50 $\mu$m \cite{Loveley-2008, Reguera-2006}. The mechanism that governs this electron transport over long-range distances is under a controversial and unresolved debate in the literature that has lasted for decades. For instance, within this debate, it was demonstrated that memristors can be fabricated using protein nanowires of \textit{Geobacter sulfurreducens}~\cite{Fu-2020} that can function at substantially lower voltages, thereby opening avenues for the construction of artificial neurons, but keeping the long-range mechanism of transport of electrons fundamentally unsolvable.

The crucial and still controversial aspect of the debate is owing to the physical mechanism associated with this long-range electron transport operating in the respiration system of living organisms, such as \textit{Geobacter sulfurreducens}, that is difficult to conciliate with the classical ET theory, as introduced in Eq.~ \ref{eq:Arrhenius-Marcus}, whether the electron transport is governed by the ballistic transmittance of the electron through the nanowires, as has been supported with experimental evidence. The experiments that support the electron transport through long-range nanometres molecular wires (referred to as the `metallic-like' mechanism) without the presence of redox states lead to a clear contradiction to classical ET theory (that supports the super-exchange mechanistic proposal). As discussed in section~\ref{sec:short-long-etransport}, classical ET theory necessarily predicts a short-range path for electrons to be transferred between $D$ (oxidation) and $A$ (reduction) states, a mechanism that is experimentally demonstrated and hence constitutes a plausible alternative for explaining the long-range electron transmittance through a sequential short-range ET between $D$ to $A$ states (see Figure~\ref{fig:D-B-A-structure}) that can operate in biological ETrCs.

It is precisely this difficulty of reconciliation between a well-established Noble Prize theory in chemistry (that supports the super-exchange ET tunneling mechanism) and ballistic transport of electrons~\citep{Imry-Landauer-1999}, that comes from a solid-state physics viewpoint, that has been the source of a hot and intense debate in the literature\cite{Strycharz-Glaven-2011, Malvankar-2011, Malvankar-2012, Strycharz-Glaven-2012}, where renowned authors are in favour of an exclusive `metallic-like' mechanism supported on microbial nanowire structures conductance~\cite{Malvankar-2011} (be it coherent~\cite{Malvankar-2011} or not~\cite{Dahl-2022}), such as pili nanofilaments, which are abundantly present in \textit{Geobacter} biofilms. 

Notably, these pili nanofilaments have conductivities in the order of $\sim$ 5 mS cm$^{-1}$, along with electrons that can be efficiently transported over distances on the order of centimeters, which is thousands of times the size of a bacterium within the biological film. Therefore, this long-range electron transmittance between $D$ and $A$ states through nanowires has been considered to be unreconciled with the classical ET theory for redox reactions. Owing to these fundamental unreconcilable circumstances is that the debate has been intense within supportive experiments (in a multitude of manners) of both (in different variances) super-exchange and `metallic-like' mechanistic scenarios.

In the next section, it will be discussed the fundamental chemistry and physical viewpoints of these long-range electron transport mechanisms that have been separately used to explain the respiration of \textit{Geobacter sulfurreducens} films.

\subsection{Chemistry or Physics of \textit{Geobacter sulfurreducens} films?} \label{sec:PhysorChem}

The debate on the long-range transport of the electrons in ETrCs is centred in a chemistry rate \textit{versus} conductance physics viewpoint of the fundamentals governing transmittance/transport of the charge carries, as was discussed in preceding sections. However, the fact is that biological processes evolved following nature's rules (or laws) that are independent from a human point of view whether these rules are based on chemical or physical concepts. The purpose of this review is to clarify that most of the difficulties in constructing a unified viewpoint between chemistry and physics of long-range transport of electrons in ETrCs are dictated by our inability to establish a conceptual correlation between ET rate and electron transport. For instance, in the same way that ET rate and electron transport concepts are used erroneously as synonymous, there are models for explaining long-range electron transport through ETrC structures in which ballistic transport and super-exchange mechanisms are used interchangeably or combined to explain different experimental results~\cite{Eshel-2020, Ru-2019}. Additionally, there are approaches in which the `metallic-like' conductance is modelled using hopping conductance at a predictable rate of 10$^{10}$ s$^{-1}$ between each hopping site, which adjusts to the viewpoint of solid-state synthetic molecular wires~\cite{Dahl-2022}.

Therefore, although the physical viewpoint is major-centred by stating that the `metallic-like' conductance mechanism of electron transport is physically supported on a ballistic conductance~\citep{Imry-Landauer-1999} concept that is well-known and experimentally tested in one-dimensional molecular structures such as molecular wires and similar structures\cite{Yang-2004, Chen-2008}, there are alternatives that consider the `metallic-like' conductance as a hopping conductance, resembling a super-exchange mechanism without the need of incorporating OmcS structures~\cite{Dahl-2022}. It is important to stress that the ballistic, `metallic-like' conductance, viewpoint of electron transport is also the foundation of the field of molecular and nanoscale electronics~\citep{Santos-2020, Bueno-2018}.

Alternative proposals to the pili-based `metallic-like' physical mechanism have been sustained on a chemistry viewpoint that is supported in the ET theory~\cite{Strycharz-Glaven-2011} and a super-exchanged mechanism, for which it is argued that electrons are transported by a succession of ET reactions among redox proteins such as cytochrome, within the outer cell membranes~\citep{Bond-2003}. This super-exchange ET mechanism of the long-range electron transport in the extra-cellular cytochrome chain of the bacteria cannot be aligned with the above viewpoint of the `metallic-like' molecular electronic physical mechanism, alluded to pili nano-filaments\cite{O'Toole-1998, Reguera-2005} present in the extracellular matrix of the bacterium, although it can be accommodated to `metallic-like' hopping conductance~\cite{Dahl-2022}.

The purpose of the present review is to demonstrate that the alignment of these mechanisms -- the super-exchange mechanism sustained on the $k$ premises of Eq.~\ref{eq:Arrhenius-Marcus} and `metallic-like' conductance supported on the conductance $G$ (be it ballistic or incoherent hopping in nature) -- can be achieved within a quantum mechanical rate interpretation of the electron transport within a quantum electrodynamics viewpoint. The conciliation of these mechanisms has previously been attempted in the literature. For instance, it was proposed an electron transport that occurs through a `stepping stone' empirical biological mechanistic viewpoint~\cite{Bond-2003, Bonanni-2013}, which proposes that there is a combined action of pili and cytochrome that sustains the long-range electron transport in Geobacter biofilms. This `stepping stone' viewpoint, albeit suitable and experimentally underpinned from a biological point-of-view, still requires a theoretical physical chemistry foundation that, for instance, would be able to connect and unify in a first-principles quantum mechanical theory $G$ (physical) and $k$ (chemical) concepts. 

Here, it is introduced, in the form of a review, a theoretical framework that establishes a reconciliation of the above $G$ and $k$ concepts~\citep{Bueno-2020-ET} and consequently conciliates super-exchange and `metallic-like' conductance mechanisms beyond the consideration of a hopping conductance mechanism~\cite{Dahl-2022} that does not incorporates the needs of redox or OmcS structures to participate in the long-range transport of electrons in complex biological structures. The theoretical basement is established through first-principles quantum and statistical mechanics and hence connects $G$ (physical) and $k$ (chemical) concepts through the quantum capacitance $C_q$~\citep{Bueno-2023-3}, as a particular setting of the electrochemical capacitance $C_{\mu}$. For more details about the meaning of $C_q$ and $C_{\mu}$, see the Support Information (SI) document, section SI.1.

This theoretical viewpoint is stated within an approach that not only encompasses the classical ET theory~\citep{Bueno-2020-ET, Bueno-2023-3}, as a particular setting of the general principles of the theory~\citep{Bueno-2023-3}, but also explains the long-range transport of electrons without disregarding the molecular (ballistic or hopping in nature) conductance of nanowire chemical or OmcS interconnected structures within ET-type of long-range electron conductance. This theory~\citep{Bueno-2023-3} is referred to as quantum rate theory and allows us to underpin a biological viewpoint of the respiration process using \textit{Geobacter sulfurreducens} as a reference of ETrC. The use of \textit{Geobacter sulfurreducens} film permits to verify the confidence of the theory directly through experiments wherein long-range electron transport has been exhaustively studied and debated in the past decade~\citep{Bond-2003, Malvankar-2012, Strycharz-Glaven-2011, Strycharz-Glaven-2012} within the above-mentioned debate between differents and, up to now, apparently unreconciled scientific viewpoints, that are intrinsically entwined in the quantum rate theory, as will be clarified in the next section.

\section{Overview of the Quantum Mechanical Rate Theory} \label{sec:QRTheory}

\begin{figure}[!t]
\centering
\includegraphics[scale=0.55] {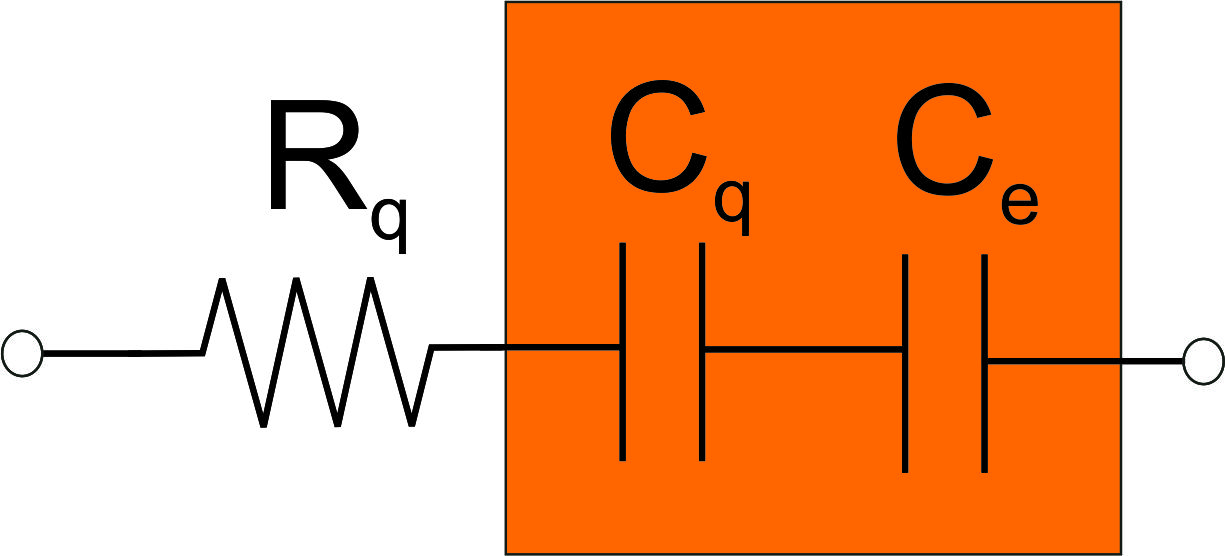}
\caption{The relativistic quantum mechanics of coherent or incoherent different electron transport modes (including the modelling of homogeneous or heterogeneous redox reactions~\citep{Bueno-2023-3}) can be described using a quantum resistive-capacitive (RC) circuit comprised of a quantum resistor $R_q = h/g_s e^2$ and of quantum capacitance $C_q$. The effect of the electrolyte is to cause a degeneracy of the energy states $e^2/C_e \sim e^2/C_q$ owing to the series combination of $C_e \sim C_q$. Therefore, as discussed in sections~\ref{sec:QRTheory} and ~\ref{sec:QED+ETr}, the electrolyte permits the long-range electron transport to follow quantum electrodynamics. This quantum electrodynamics of ET and transport of electrons \textit{per se} reconciles the debate of ET \textit{versus} coherent or incoherent conductance analyses because of the entwined relations between these two concepts through $C_q$ within a degenerate state of energy $E = e^2/C_q$ provided by the role played by the electrolyte in the electron transport and dynamics.}
\label{fig:QRC-circuit}
\end{figure}

This section aims to make a concise introduction of the quantum mechanical rate concept (Eq.~\ref{eq:G/C} below) to demonstrate that it encompasses `metallic-like' (ballistic or incoherent hopping) conductance as well as the super-exchange (with redox or intra-molecular ET dynamics) mechanistic viewpoint through the concomitant consideration of quantum conductance and ET rate dynamics as essential and entwined constitutive components of the theory. The fundamental entwining of quantum conductance and ET concepts are intrinsically part of an understanding of how the quantum rate concept reconciles the debate within different viewpoints containing `metallic-like' (coherent or incoherent) conductance and super-exchange (incorporating redox ET reactions or intra-molecular ET) mechanisms which are contained in the context of the long-range electron transport of biological respiration chains, as introduced in previous sections.

To demonstrate how the intense and heated debate on the long-range electron transport phenomenon can be harmonized using first-principle quantum rate fundamentals that encompasses three ($G$, $k$ and $C_\mu$) different concepts~\citep{Bueno-2020-ET,Bueno-book-2018}, it must be noted that the quantum rate constant $k$ is simply established as the ratio between the quantum of conductance $G(\mu) = G_0 \sum_{n=1}^{N} T_n(\mu)$ and the electrochemical capacitance $C_{\mu}$, such as \cite{Bueno-2020, Bueno-2018, Bueno-book-2018}

\begin{equation}
 \label{eq:G/C}
	k = \frac{G(\mu)}{C_{\mu}} = G_0 \sum_{n=1}^{N} T_n(\mu) \left( \frac{1}{C_e} + \frac{1}{C_q} \right), 
\end{equation}

\noindent where $1/C_{\mu} = 1/C_e + 1/C_q$ term is identified as the series combination of geometric $C_e$ and quantum $C_q$ capacitances within $C_{\mu}$ as an equivalent electrochemical capacitance~\cite{Bueno-2019} (see more details in SI.1). Note that $G(\mu)$ is well-known as the Landauer conductance~\cite{Landauer-1957}, where $G_0 = g_s e^2/h$ is a constant referred to as the conductance quantum, possessing a value of $\sim$ 77.5 $\mu$S, $e$ refers to the elementary charge and $g_s = 2$ states for the spin degeneracy of the electron. Note that the consideration of the spin dynamics in the quantum rate model is important and agrees with theoretical studies devoted to the consideration of the spin-dependent characteristics of long-range electron transport in multi-heme bacterial nanowires~\citep{Livernois-2023}.

As can be noted in Eq.~\ref{eq:G/C}~\citep{Bueno-2020-ET, Bueno-2023-3, Bueno-book-2018}, the quantum conductance $G$ and ET rate constant $k$ are related through the electrochemical capacitance $C_\mu$ concept. Note that the $k$ rate concept can be simply represented in terms of experimentally measurable equivalent circuit parameters, as noted in Figure~\ref{fig:QRC-circuit}. This section will demonstrate the relationship between these three concepts and how the quantum rate amendment of electron transport conforms with a relativistic quantum electrodynamics viewpoint of the process. The intrinsic relativistic quantum dynamics of the electron transport within the quantum rate theory corresponds to a dynamics of transport of energy that is long-range \textit{per se} providing there is an electrolyte medium (intrinsic of biology processes). The role of the electrolyte medium is to permit a suitable electric field screening to transmit the electron at long distances which is the intrinsic consequence of low-frequency electrodynamics; hence implying the lowest expense of energy for the transmittance of the electron at long distances.

It has been demonstrated that the quantum rate theory, as summarized in Eq.~\ref{eq:G/C}, encompasses Marcus's ET theory~\citep{Bueno-2020-ET} as a particular setting of the quantum rate concept whenever the statistical mechanics (see section~\ref{sec:QED-SMech} for more details) is considered into Eq.~\ref{eq:G/C}, straightforwardly leading to Eq.~\ref{eq:Arrhenius-Marcus}. The proof that Eq.~\ref{eq:Arrhenius-Marcus} is a particular setting of Eq.~\ref{eq:G/C}~\citep{Bueno-2020-ET, Bueno-2023-3} enlightens the fact that quantum rate theory is able of dealing with electrochemical reaction processes in different settings, including heterogeneous ET reactions, besides the homogeneous ET setting predicted by Eq.~\ref{eq:Arrhenius-Marcus}, as originally proposed by Marcus~\citep{Marcus-1964}.

In other words, the quantum rate theory allows us to model electron exchanged or transferred between any $D$ and $A$ states, where redox ET reactions~\citep{Bueno-2020,Bueno-2023-3} or intermolecular ET~\citep{Bueno-2023-ICT} mechanisms are established solely as particular settings of the general quantum rate concept. In both ET reactions~\citep{Bueno-2023-3} and intra-molecular ET~\citep{Bueno-2023-ICT} situations, the phenomenon obeys quantum mechanical rules that cannot be correctly modeled using Schr\"ordinger's non-relativistic quantum-wave methods but can be using relativistic quantum electrodynamics~\citep{Dirac-1928}. The theory defines a fundamental quantum rate $\nu$ principle based on the ratio between the reciprocal of the von Klitzing constant $R_k = h/e^2$ and the quantum (or chemical) capacitance, such as~\citep{Bueno-2023-3}

\begin{equation}
 \label{eq:nu}
	\nu = \frac{e^2}{hC_q},
\end{equation}

\noindent where by noting that $E = e^2/C_q$ is the energy intrinsically associated with the electronic structure (see more details in SI document), it straightforwardly leads to the Planck-Einstein relationship, i.e.

\begin{equation}
 \label{eq:Planck-Einstein}
	E = h\nu = \hbar \textbf{c}_* \cdot \textbf{k},
\end{equation}

\noindent which follows a linear relationship dispersion between the energy $E$ and wave-vector\footnote{Note that \textbf{k} in bold refers to the wave-vector and its magnitude will be referred to as |\textbf{k}| to avoid notation issues with the meaning of $k$ that, in previous works~\citep{Bueno-2023-3,Sanchez-2022-1} and Eq.~\ref{eq:G/C}, was referred to as the electron transfer rate constant.} $\textbf{k}$, where $\textbf{c}_*$ is the Fermi velocity and $\hbar$ is the Planck constant $h$ divided by $2\pi$. Hence, the quantum rate principle within Eq.~\ref{eq:nu} predicts relativistic dynamics that comply with Dirac~\citep{Dirac-1928} instead of with the non-relativistic Schr\"ordinger equation, which by invoking the De Broglie relationship, which states that the momentum $\textbf{p} = \hbar \textbf{k}$ is directly related to the $\textbf{k}$, can be verified and hence Eq.~\ref{eq:Planck-Einstein} turns into $E = \textbf{p} \cdot \textbf{c}_*$, demonstrating its intrinsic relativistic character. Observe that the relativistic quantum electrodynamics of Eq.~\ref{eq:Planck-Einstein} and consequently of Eq.~\ref{eq:G/C}, as a particular setting of phenomenon stated in Eq.~\ref{eq:Planck-Einstein}, implies that the electron can behave as a wave (or a massless particle) similarly to the electrodynamics stated in two-dimensional compounds such as graphene~\citep{Bueno-2022}. Therefore, it is not a coincidence that the quantum rate theory applies to explain redox reactions~\citep{Bueno-2023-3}, intra-molecular charge transfer~\citep{Bueno-2023-ICT} and the quantum electrodynamics of graphene~\citep{Bueno-2022} embedded in an electrolyte environment.

\textit{De facto}, it has been experimentally demonstrated~\citep{Alarcon-2021, Bueno-2023-3, Sanchez-2022-2} that $E = e^2/C_q$ is a degenerated state of energy for experiments conducted in an electrolyte environmental medium, such as that effectively $E$ of Eq.~\ref{eq:G/C} is stated as $g_s g_e (e^2/C_q)$, where $g_s$ is the electron spin degeneracy and $g_e$ is the energy degeneracy state associated with the electric-field screening effect of the electrolyte over the molecular states. Note that the degenerate state of Eq.~\ref{eq:Planck-Einstein}, that is $g_s g_e (e^2/hC_q)$, is achieved by considering a specific setting of Eq.~\ref{eq:G/C}. The specific situation of Eq.~\ref{eq:G/C} that leads to $k$ to equate to $\nu = g_s g_e E/h$, that is, to a degenerate state of Eq.~\ref{eq:Planck-Einstein}, is simply taken by assuming that $C_e \sim C_q$ and an additional adiabatic condition in which $\sum_{n=1}^{N}T_{n}\left( \mu \right)$ must equate to unit~\cite{Alarcon-2021, Bueno-2023-3} in Eq.~\ref{eq:G/C}. Nonetheless, non-adiabatic or any other electronic scattering condition can be considered simply by noting that $\nu = \left( g_s g_e E/h \right) \sum_{n=1}^{N}T_{n}\left( \mu \right)$, where different electron scattering settings can be formulated by the appropriate use of the transmission matrix $\sum_{n=1}^{N}T_{n}\left( \mu \right)$ at various conditions to be established according to the biological system to be modelled.

There are two equivalent interpretations of $g_e$ degeneracy. The first is based on the existence of resonant electric currents, i.e., there are time-dependent (displacement) electric currents within the system. This type of resonant electric current (ambivalent - see SI.3 for more details) is owing to the existence of two charge carriers (electrons and holes) that promote a net current denoted here as $i_0$. The origin of this displacement electric current is owing to the role played by the electrolyte that allows the superimposition of the electrostatic $C_e$ and quantum capacitive $C_q$ modes of charging the system. In this situation, the elementary charge $e$ is subjected to an equivalent electric potential, i.e., $e/C_e \sim e/C_q$, conducting locally to $C_e \sim C_q$.

Hence, the equivalent $C_\mu$ capacitance of the junction is $1/C_\mu = 2/C_q$, with an energy degeneracy of $e^2/C_\mu = 2e^2/C_q$, where 2 is accounted as $g_e$ in the formulation of $\nu = g_e G_0/C_q$ concept (see SI.2). This energy degeneracy is equivalent to the previous electrical current degeneracy consideration of the origin of $g_e$ because $1/C_\mu = 2/C_q$ implies $C_q = 2C_{\mu}$. Noting that the capacitance of the system is a result of the equivalent contribution of $C_e$ and $C_q$, there is an electric current degeneracy for charging the quantum capacitive states that are proportional to $C_\mu$ with an electric current of $i_0 = C_{\mu}s = (1/2)(C_q)s$, which is equivalent to $2 i_0 = C_q s$, where $s = dV/dt$ is the time potential perturbation (scan rate) imposed to the system. Noteworthy that a time-dependent perturbation $s$ (or equivalent) is required to investigate the energy $E = e^2/C_q$ level, electronically coupled to the electrode states. Accordingly, it is owing to the equivalence of $e^2/C_e$ and $e^2/C_q$ energy states that there is an energy degeneracy that inherently permits two electric currents contributing to the net (electrons and holes) current $i_0$ of the system.

In Eq.~\ref{eq:G/C}, $G(\mu) = G_0 \sum_{n=1}^{N} T_n(\mu)$ states for the conductance quantum meaning in its manifold representation, signifying that multiple individual $n$ quantum conductance channels for transport of the electron contribute to the total quantum conductance $G$. In this manifold representation of $G$, $T_n(\mu)$ states represent the transmission probabilities across each channel at a given chemical potential state $\mu$. Hence $\sum_{n=1}^{N} T_n(\mu)$ is a statistical component of $G$ known as transmission matrix that permits modeling any type of electron scattering mechanism. For instance, through this scattering approach coherent or incoherent tunneling as well as coherent or incoherent hopping, to mention but a few, can be modeled within the quantum rate theory. For a particular ideal setting of a perfect electron transmittance, the transmittance probability equates to the total number of channels $N$. In other words, for a perfect mode of transmittance of electrons in multiple ideal channels, it implies that $G = G_0 N$ and that $G$ is equivalent to $G_0$ for a single individual channel perfect mode of electron transmittance, where thus $N$ is settled as unity in $G = G_0 N$, physically representing an adiabatic mode of transport of a single electron.

Accordingly, adiabatic (which is case of both intra-molecular charge transfer~\citep{Bueno-2023-ICT} and electrodynamics of graphene~\citep{Bueno-2022}) or non-adiabatic (ET reactions~\citep{Bueno-2023-3}) modes of transport can be modeled by Eq.~\ref{eq:G/C} through $\sum_{n=1}^{N} T_n(\mu)$ statistical term. For instance, consider an adiabatic single electron transport to be modeled, constituting a situation in which $\sum_{n=1}^{N} T_n(\mu)$ is unity because $N$ is unity and the transmittance is ideal (adiabatic), hence allowing to settle Eq.~\ref{eq:G/C} simply as $k = G_0/C_\mu$ and the degeneracy state of Eq.~\ref{eq:Planck-Einstein} is achieved simply by considering $C_e \sim C_q$ in Eq.~\ref{eq:G/C}, as discussed in more detail in the SI document.

In summary, analysing Eq.~\ref{eq:G/C}, under the situation of modeling an adiabatic single electron transport in an electrolyte medium, supported in experimental evidence that accounts for the key role of the electrolyte environment~\citep{Pinzon-2022}, where it is noted a superposition of $C_e$ and $C_q$ capacitive states owing to an appropriate electric-field screening mechanism of the electrolyte over $D$ and $A$ states~\citep{Bueno-2023-3}, it follows that Eq.~\ref{eq:G/C} turns into~\citep{Alarcon-2021}

\begin{equation}
\label{eq:G/C-degeneracy}
	k = g_s g_e \left( \frac{e^2}{hC_q} \right) = g_e \left( \frac{G_0}{C_q} \right) = g_s g_e \left( \frac{E}{h} \right). 
\end{equation}

Eq.~\ref{eq:G/C-degeneracy} was successfully verified by experiments~\citep{Alarcon-2021,Bueno-2023-3} and theoretically demonstrated to comply with Laviron's approach~\citep{Laviron-1979} for a diffusionless ET dynamics~\citep{Alarcon-2021}, intra-molecular charge transfer in molecular resonant junctions~\citep{Bueno-2023-ICT} and electrodynamics of graphene~\citep{Bueno-2022}. The origin of $g_e$ (that in the particular situation of redox reaction is referred to $g_r$~\citep{Alarcon-2021}) conforms with an electrical current degeneracy consideration where it can be noted that $g_e$ is related to a particular electric-field screening that implies that $1/C_\mu = g_e/C_q$, as discussed in section~\ref{sec:QRTheory}, where it was demonstrated that $C_q = 2C_{\mu}$. For the particular case of redox reactions, it inherently conducts to two (anodic and cathodic) electrochemical currents in redox reactions~\citep{Bueno-2023-3}, for instance, contributing to the exchange net current $i_0$ of the interface~\citep{Bueno-2023-3}, implying that the faradaic electrochemical current is ambivalent in redox reaction dynamics~\citep{Bueno-2023-3}.

This ambivalent electric current phenomenon (see SI. 3) is equivalently observed in resonant molecular junctions~\citep{Bueno-2023-ICT} and graphene~\citep{Bueno-2022}, where the quantum rate theory applies successfully despite there be no redox reaction. In other words, Eq.~\ref{eq:G/C-degeneracy} implies only the consideration of additional $g_e$ degeneracy besides $g_s$ in the electron transport rate mechanism stated by Eq.~\ref{eq:G/C}, wherein the specific case of an adiabatic transport of a single electron, the meaning is that $G$, in Eq.~\ref{eq:G/C}, is equivalent to $G_0$. This boundary condition implies simply considering a single adiabatic electron transport ideal situation in which $\sum_{n=1}^{N}T_{n}\left( \mu \right) = N\exp \left(-\beta L \right) = N\kappa \sim 1$, or a non-adiabatic situation in which $k = \kappa g_e G_0/C_q$, with $\kappa$ explicitly considered in the expression of $k$ to take into account the allowed multitude of electron scattering (including coherent and incoherent tunneling and hopping) possibilities besides the ballistic transport, as a particular setting, as already noted previously. For instance, $\kappa$ in $k = \kappa g_e G_0/C_q$ implies the consideration of a non-adiabatic electron transport which is modeled as taking $\kappa$ lower than unity, but higher than null for electron transmittance to occur. As will be better noted in section~\ref{sec:Theor+Exper}, no matter the type of electron scattering mechanism adopted to model the quantum rate, an equivalent circuit of the type shown Eq.~\ref{fig:QRC-circuit} is experimentally measurable, which turns the quantum rate mode easily applicable to different complex situations, including biological processes.

In adiabatic~\cite{Bueno-2023-ICT, Bueno-2022} or non-adiabatic~\cite{Bueno-2023-3} situations, it has been experimentally demonstrated~\cite{Bueno-2023-ICT, Bueno-2023-3, Bueno-2022} that the charge transfer resistance is equivalent to the resistance quantum $R_q = 1/G_0 = h/g_se^2 \sim$ 12.9 k$\Omega$, which experimentally validates the quantum rate theory~\citep{Sanchez-2022-2} in different modes of transport of electrons. The demonstration of a quantum limit value of $R_q$ for the transport of electrons is achievable because $\kappa$ (for non-adiabatic processes) can be experimentally measured as well as the number of channel $N$ that is obtained from the measurement of $C_q$~\citep{Sanchez-2022-2}. Therefore, besides quantum rate theory~\citep{Bueno-2020-ET, Bueno-2023-3, Bueno-book-2018} encompasses other electron transport models and approaches (such as homogeneous ET) developed by Marcus~\citep{Marcus-1985}, the advantage of this theory is the easier experimental test and applicability owing to all the fundamental parameters of the theory are measurable.

The next section will discuss the consequences of the quantum electrodynamics meaning of Eq.~\ref{eq:G/C-degeneracy}, an intrinsic electrodynamics characteristic of the quantum rate concept that additionally and, self-consistently, helps us to understand the long-range electron transport in respiration chains.

\section{The Quantum Electrodynamics of Electron Transport} \label{sec:QED+ETr}

\begin{figure}[!t]
\centering
\includegraphics[scale=0.44] {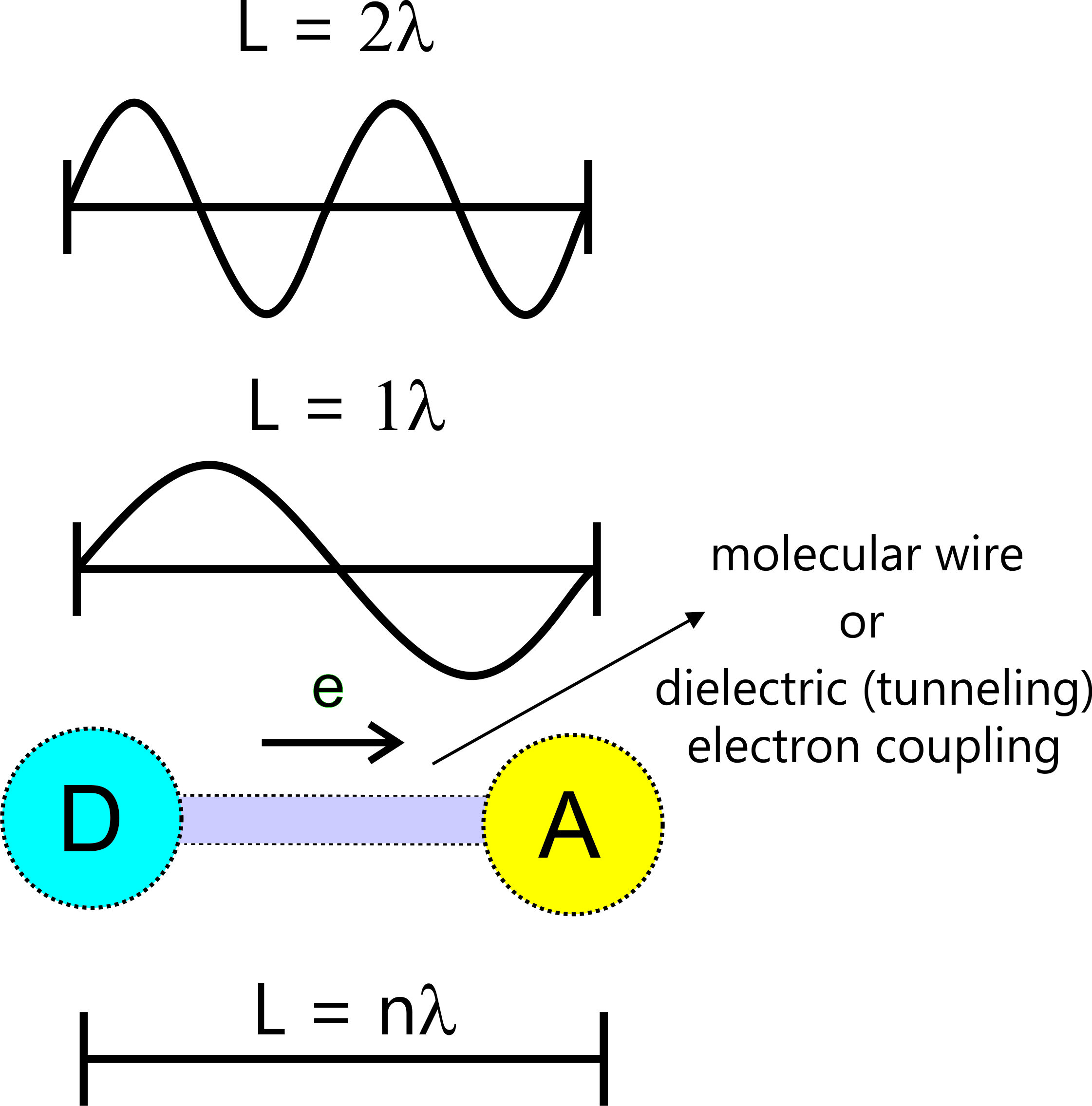}
\caption{According to the quantum rate theory, differences between the transport of electrons through redox short range tunneling mechanism and `metallic-like' (incoherent or incoherent) mechanisms are only a matter of configuration of the bridge (through space or molecular), i.e. $k = \kappa g_s g_e (e^2/hC_q) = (c_*/L)$ or equivalently $h\nu = \hbar \textbf{c}_* \cdot \textbf{k} =\kappa g_s g_e E$, where $\kappa$ (electron coupling or any type of scattering statistics modelled by the transmission matrix $\sum_{n=1}^{N} T_n(\mu)$ component of the quantum rate model) can acquire different statistics according to the dominant type of electron transport to be modelled. Different models of electron transport can be accommodated within the quantum rate theory considering a long-range $L$ within a low-frequency regime that permits $\lambda$ to be large enough to allow electrons to be transported through millimetres lengths within a ballistic, super-exchange or a combination of these modes of electron transport. For instance, as demonstrated in section~\ref{sec:Theor+Exper}, $D$ and $A$ can be, for instance, a redox pair separated by a dielectric (in the case of molecular films~\citep{Bueno-2023-3}) medium with electron transport operating via ET tunneling mechanism or they can be states separated by a molecular wire bridge~\citep{Bueno-2023-ICT} that follows coherent (ballistic) or incoherent (sequential hopping) mechanisms of transport.}
\label{fig:ET+ballistic}
\end{figure}

In the previous section, it was demonstrated that Eq.~\ref{eq:G/C-degeneracy} implies relativistic quantum electrodynamics~\citep{Bueno-2023-3}, such as that $e^2/hC_q$ corresponds to a $\nu = e^2/hC_q$ frequency that complies with the Planck-Einstein relationship, as stated in Eq.~\ref{eq:Planck-Einstein}. Hence, owing to $k = g_s g_e \nu$, the difference between the $\nu$ and $k$ frequencies is only numerical and accounted by $g_s$ and $g_e$ energy degeneracies. Accordingly, it is confirmed that the quantum rate concept that governs ET reactions~\citep{Bueno-2023-3}, intra-molecular charge transfer~\citep{Bueno-2023-ICT}, and graphene electrodynamics~\citep{Bueno-2022} follows a linear relationship dispersion between $E$ and $\textbf{k}$ and thus Eq.~\ref{eq:G/C-degeneracy} conforms with relativistic quantum electrodynamics, as illustratively depicted in Figure~\ref{fig:ET+ballistic}.

The previous analysis implies that the quantum rate theory obeys quantum electrodynamics for different models of electron transport in an electrolyte environment and room temperature, with electrons following Dirac~\citep{Dirac-1928} relativistic wave mechanics. However, this does not imply that electrons are travelling close to the light's velocity (which is generally the common sense for relativistic dynamics), but it does means that the electron follows a photonic dynamics predicated by Eq.~\ref{eq:Planck-Einstein} in its relativistic character, i.e. a wave behaviour that is referred to a massless particle dynamics. Photons are massless particles with integer spin of $+1$ or $-1$, whereas electrons are particles with spin of $+1/2$ or $-1/2$ called Fermions. Electrons with null rest mass are specifically named massless Fermionic particles. The massless Fermionic relativistic electrodynamics of the electron is given by considering the following relativistic relation $E^2 = p^2c_*^2 + m_0^2 c_*^4$, where $E$ is the total energy and $m_0$ is the rest mass of the electron. A massless Fermionic behaviour of the electron is when its rest mass $m_0$ is null, implying that $E = pc_*$ and thus obeying the relativistic electrodynamics predicted by Eq.~\ref{eq:Planck-Einstein}, \textit{quod erat demonstrandum}.

Accordingly, this relativistic dynamics of electron transport governed by $E = h \nu = \hbar \textbf{c}_* \cdot \textbf{k}$ simply equation resembles the transmittance of electromagnetic waves (or photonic dynamics) through space. In the case of the transport of electrons, this intrinsically requires a time-induced charge-fluctuation analysis that is based on considering individual $n$ quantum channels, each of length $l$ (for individual ET short-range steps) or $L$ (for long-range electron transport) with a given density of $(dn/dE)$, with a Fermi velocity $c_*$ for electrons to travel in the channels, permitting to model changes in the electrochemical faradaic or coherent and incoherent electric currents as~\citep{Bueno-2023-3}

\begin{equation}
 \label{eq:electric-current-Qrate}
	\delta i = -e \left( \frac{c_*}{L} \right) \delta \mu \left( \frac{dn}{dE} \right),
\end{equation}

\noindent from which $G$ is obtained by substituting $\delta \mu = -e \delta V$ in Eq.~\ref{eq:electric-current-Qrate} and further rearranging leads to

\begin{equation}
 \label{eq:G-v_F-DOS}
	G = \frac{\delta i}{\delta V}  = e^2 \left( \frac{c_*}{L} \right) \left( \frac{dn}{dE} \right),
\end{equation}

\noindent from which, by noting the definition of the DOS for a perfect quantum channel as $(dn/dE) = g_s L/c_*h$ and substituting it in Eq.~\ref{eq:G-v_F-DOS}, it is obtained that $G = g_s e^2/h = G_0$. Noting that $C_q = e^2(dn/dE)$, Eq.~\ref{eq:G-v_F-DOS} can be also simplified to as

\begin{equation}
 \label{eq:G-v_F}
	\frac{G}{C_q} = \left( \frac{c_*}{L} \right),
\end{equation}

\noindent which is equivalent to Eq.~\ref{eq:G/C-degeneracy} whenever $g_e$ is considered in the definition of $\nu = g_s e^2/hC_q = G_0/C_q$ rate. This is an expected result based on the quantum rate description of a single and ideal ET or electron transport step with specific energy degeneracies, as discussed in section~\ref{sec:QRTheory}.

For the sake of simplicity in the physical analysis of the problem, let us consider adiabatic electron transport where $\kappa$ is unity. In this case, the meaning of $c_*/L$ is settled by noting that $L = n\lambda$, where $n$ is the number of quantum state modes within $L$ for the quantum transmittance, i.e. $nc_*/\lambda$. 

Implicit in the analysis of $G/C_q = c_*/L$ is that if $G$ is normalized by $N$ (the total number of channels $n$), which is obtained in the Fermi level of ET reactions or electron transport in molecular wires, the corresponding situation is $G_0/C_q = c_*/\lambda$. The latter is owing to $G_0 = G/N$ for the adiabatic ET or electron transport situations. This particular setting is owing to $C_q = e^2(dn/dE)$ (see SI.3) which, as theoretically and experimentally discussed in previous works~\citep{Alarcon-2021,Bueno-2020-ET,Bueno-2023-3,Bueno-book-2018} (see also SI document), corresponds to a $C_q$ that is proportional to the density-of-states DOS $(dn/dE)$ over conductive electrodes, from which $N$ can be measured.

In the next section, it will be demonstrated that the relativistic quantum electrodynamics can be considered despite of the influence of the temperature over the energy states of the system, which is of particular importance to validate the quantum rate approach for room temperature electron transport within biological chains.

\section{Statistical Mechanical of the Quantum (Rate) Energy States} \label{sec:QED-SMech}

Noteworthy is that $C_q$ contains all the required information for the analysis of ET reactions~\citep{Miranda-2016, Miranda-2019} and coherent or incoherent electron transport and the energy associated with $C_q$, that is $E = e^2/C_q = h \nu$, has a meaning that depends on the temperature in which it is stated, i.e. with or without (at the zero-temperature limit) statistical mechanics consideration. Without a thermal dependent context $E = e^2/C_q = h \nu$ is not consistent with biological processes that occur mainly at room temperature. The formulation of electron transport at the zero-temperature has only a theoretical and didactic purpose disconnected from biological reality.

In other words, the temperature dependence required for a room temperature analysis of the ET dynamics of electrochemical reactions or coherent/incoherent electron transport can be only evaluated considering appropriate statistical mechanics that provide significance for the energy distribution but still keeps quantum electrodynamics characteristics. Accordingly, the consideration of the thermodynamics conducts to distribution and thermal dependence of $E = e^2/C_q$ that is appropriately stated using the grand canonical ensemble presumption~\citep{Bueno-2023-3}, from which it arises that $E = e^2/C_q = k_BT/N \left[f(1-f) \right]^{-1}$, where $f = \left(1 + \exp  \left(\Delta E / k_{B}T \right) \right)^{-1}$ is the Fermi-Dirac distribution function, $k_B$ is the Boltzmann constant, $T$ is the absolute temperature and $\Delta E = \mu - E_F = -eV'$ is an energy difference established concerning Fermi level $E_F$ of the system (see also SI. 2). Within statistical mechanics considerations, Eq.~\ref{eq:G/C} can be rewritten as~\citep{Bueno-2023-3}

\begin{equation}
 \label{eq:k-finiteT}
	k = \frac{G}{C_q} = G_0 \left( \frac{k_BT}{e^2N} \right) \left[f(1-f) \right]^{-1} \sum_{n=1}^{N}T_{n}\left( \mu \right).
\end{equation}

\begin{figure}[!t]
\centering
\includegraphics[scale=0.44] {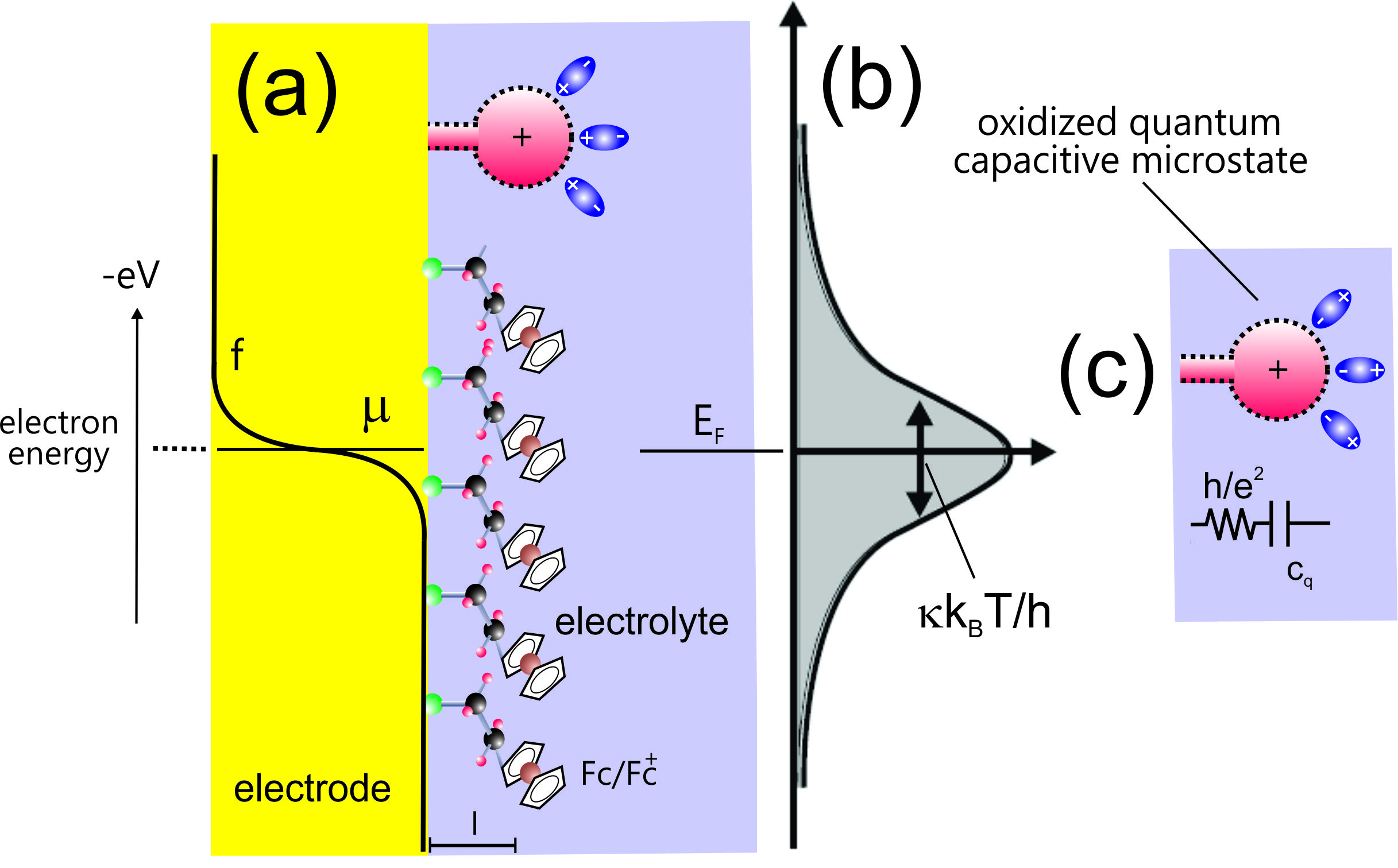}
\caption{(a) Illustration of a redox-active molecular film of thickness $l$ (lower than 2 nm) assembled on an electrode. The electron states of the electrode are in equilibrium with the electrons in the film through the occupancy probability function $f$, which spreads as shown in (b). (c) Each molecule in the film behaves as a quantum capacitive point contact electrically modelled as a quantum RC circuit with individual quantum conductive e$^2$/h and capacitive $C_q$ elements, forming an ensemble that is experimentally measured. The difference between this molecular film and that shown in Figure~\ref{fig:biofilm} for the biological film is only attained by the statistical mechanics occupancy of the charges within $C_q$ term of the circuit. The statistical occupancy of the energy states $E = e^2/C_q$ depends on the thickness of the films and the properties of the electron coupling $\kappa$ between $C_q$ states within the length of these films.}
\label{fig:molecular-film}
\end{figure}

It can be observed that Eq.~\ref{eq:k-finiteT}, besides comprising any type of electron scattering processes through $\sum_{n=1}^{N}T_{n}\left( \mu \right)$ term, also incorporates the required statistical mechanic's considerations of the quantum mechanics of the rate. Straightforward related to Eq.~\ref{eq:k-finiteT} is thermal dependence of $C_q$ that is accounted as~\citep{Bueno-book-2018} (see also SI.3)

\begin{equation}
 \label{eq:Cq-thermal}
	C_{q} = \left( \frac{e^2N}{k_{B}T} \right) \left[ f(1-f) \right],
\end{equation}

\noindent from which it can be demonstrated that quantum electrodynamics within Eq.~\ref{eq:G-v_F} remains quantized independently of the thermal energy (or finite temperature) contribution, as illustrated in Figure~\ref{fig:molecular-film}\textit{b}. 

The quantum electrodynamics at room temperature is solely possible owing to the energy degeneracy considered within Eq.~\ref{eq:G/C-degeneracy}, as will be now demonstrated. Let us start by showing that within Eq.~\ref{eq:G-v_F} is also the meaning of $i_0$ established as a function of $c_*$ from which it should be observed that any definition of resistance must be correlated with the velocity of the carriers. Noting from previous works~\citep{Bueno-2023-3,Sanchez-2022-2} that $G_0 = (e/k_BT) i_0$, it implies a thermal voltage of $k_BT/e = i_0/G_0$ at the Fermi level, which for considering a single electron transmission process, the total modes $N$ of transmittance was equated to the unit. From Eq.~\ref{eq:Cq-thermal}, $k_BT/e$ is equivalently measured as a function of $C_q$ as $e/4C_q$, also by considering $f = 1/2$ at the Fermi level and $N$ as the unit. Hence, by combining the thermal voltages obtained from the meaning of $G$ and $C_q$, it is possible to demonstrate, taking into account the $g_e$ energy state degeneracy, that $i_0$ is~\citep{Bueno-2023-3} (see also SI.3 an alternative and more detailed way of deriving this key relationship)

\begin{equation}
 \label{eq:i0-general}
	i_0 = \left( \frac{g_e e}{4} \right) \left( \frac{G_0}{C_q} \right) = e \left( \frac{e^2}{hC_q} \right),
\end{equation}

\noindent where the value of 4, which comes from the statistical mechanical occupancy of the states, is compensated by the product of $g_s = 2$ and $g_e = 2$, \textit{quod erat demonstrandum}. The case of a single adiabatic ET implies that $L = \lambda$ and taking Eq.~\ref{eq:G-v_F} for this situation, Eq.~\ref{eq:i0-general} can be rearranged as~\citep{Bueno-2023-3}

\begin{equation}
 \label{eq:i0}
	 \nu = \left( \frac{e^2}{hC_q} \right) = \left( \frac{c_*}{\lambda} \right) = \frac{i_0}{e},
\end{equation}

\noindent which not only establishes a direct correlation between the exchange (or ambipolar) current $i_0$ and $c_*$, as is required for an appropriate physical description of the meaning of charge transfer resistance $R_{ct} = 1/G_0$ of a single electron in the field of electrochemistry~\citep{Sanchez-2022-1, Bueno-2023-3}, but Eq.~\ref{eq:i0} also establish the meaning of $i_0$ directly as a function of $\nu$ through the elementary charge of the electron $e$. Noteworthy is that $i_0 = e\nu$, where $i_0 \propto c_*$ is established as $e/\lambda$, which is the electron density in the quantum channel. 

The latter proportionality is intrinsic to the quantum electrodynamics nature of redox reactions~\citep{Bueno-2023-3}, intra-molecular charge transfer in molecular junctions~\citep{Bueno-2023-ICT} and graphene~\citep{Bueno-2022}, with quantum conductance $G$ intrinsically contained in the phenomenon of the quantum rate. This nature implies that $i_0$ of a single electron is not only ambipolar (see SI. 3) but also has a time-dependent meaning that implies electric current dynamics with a net value at the Fermi level within equivalent amounts of electron and hole carriers flowing in opposite directions. The physical meaning of the latter statement for redox reactions~\citep{Sanchez-2022-1} is that there is the transport of electron (reduction) and hole (oxidation) carriers flowing in opposite directions at the equilibrium of the reaction whereas the rate of the reaction is purely defined in terms of chemical kinetics (name exchange current), where the meaning of the ambivalence of electric current is generally disregarded. In other words, the oxidation and reduction electric currents are of the same magnitude but in the opposite direction for electrochemical reaction at equilibrium, implying the presence of an exchange current of $i_0$, which in terms of physical terminology is ambipolar and related to $g_e$, as discussed in section~\ref{sec:QRTheory}.  

This quantum mechanical character of the transport of energy through electric ambipolar current $i_0$ has been experimentally demonstrated in redox reactions~\citep{Bueno-2023-3, Alarcon-2021, Sanchez-2022-2}, resonant molecular junctions~\citep{Bueno-2023-ICT} and graphene~\citep{Bueno-2022} and, in all of the cases, it follows a quantum RC dynamics where $\nu = 1/R_qC_q$. Particularly in the case of redox reactions, the quantum capacitive states attached to a metallic interface can mediate the redox reactions if redox states are energetically aligned~\citep{Sanchez-2022-1} and permit a significant improvement in the electron transport between the electrode and free redox species in solution (long-range transport of electrons in a diffusionless kinetic dynamics), which cannot be achieved by the direct contact of metal with free redox states in the solution/electrolyte bulk phase~\citep{Sanchez-2022-1}. It was shown that the mediation of ET reactions through quantum capacitive states can effectively improve the electron transfer to a quantum efficiency whether $C_q$ states are intermediating the transport~\citep{Sanchez-2022-1}, permitting the transport of electrons at long distances in diffusionless (without mass control) kinetic mode. 

The quantum efficiency of electron transport is obtained independently of a non-adiabatic bridge within an electron coupling of $\kappa \sim \exp \left( -\beta \lambda \right)$ for the transfer of a single electron (or a single channel mode) through multiple barriers of length $l$. Furthermore, it has been shown that the resistance that obeys quantum rules (or a quantum limit with a $R_q = 1/G_0$ magnitude) in the interface is the total series resistance that includes contact- and solution-phase resistances.

In the next section, it will be analyzed experiments conducted in nanometre and \textit{Geobacter sulfurreducens} micrometer scale films, comparatively. It will be shown that these two different experimental set-ups can be equivalently modeled by Eq.~\ref{eq:k-finiteT} considering Fermionic or Boltzmannian statistics of electron occupancy in these films. In other words, it will be demonstrated that both films follow equivalent relativistic quantum electrodynamics at room temperature solely differing by the statistical mechanical settings of the energy distribution within the length of these films.

\section{Theoretical and Experimental Evidences in Support of the Quantum Rate Model} \label{sec:Theor+Exper}

Observe that, according to the analysis conducted in section~\ref{sec:QED-SMech}, the quantum relativistic character of Eq.~\ref{eq:k-finiteT} implies a time-dependent dynamics in which the electron transport remains quantized regardless of the transmission mode~\cite{Yacoby-1996}, as illustrated in Figure~\ref{fig:ET+ballistic}. This important outcome of the analysis is owing to Landauer~\citep{Landauer-1957} formalism of quantum conductance that combined with the characteristics of $C_q$ intrinsically conduct to an electron transport regime that follows a quantum RC electrodynamics. Hence, this characteristic quantum RC circuit follows relativistic dynamics that can be easily investigated experimentally using impedance-derived capacitance spectroscopy, in which a complex capacitance function can be obtained as~\citep{Bueno-2023-3}

\begin{equation}
 \label{eq:Complex-Cq}
	C^{*}(\omega) = \frac{C_{q,0}}{1 + j\omega \tau} \sim C_{q,0} \left(1 - j\omega \tau \right),
\end{equation}

\noindent where the term $C_{q,0}$ corresponds to the equilibrium capacitance, which is obtained for $\omega \rightarrow 0$, where the imaginary component of the complex capacitance is negligible. $\tau = R_q C_q$, in Eq.~\ref{eq:Complex-Cq}, corresponds to the characteristic quantum RC relaxation time where $R_q = 1/G$, summarized in terms of circuit elements in Figure~\ref{fig:QRC-circuit}. Accordingly, the rate constant of the electron transport is obtained simply as $k = 1/\tau$.

\begin{figure}[ht]
\centering
\includegraphics[scale=0.49] {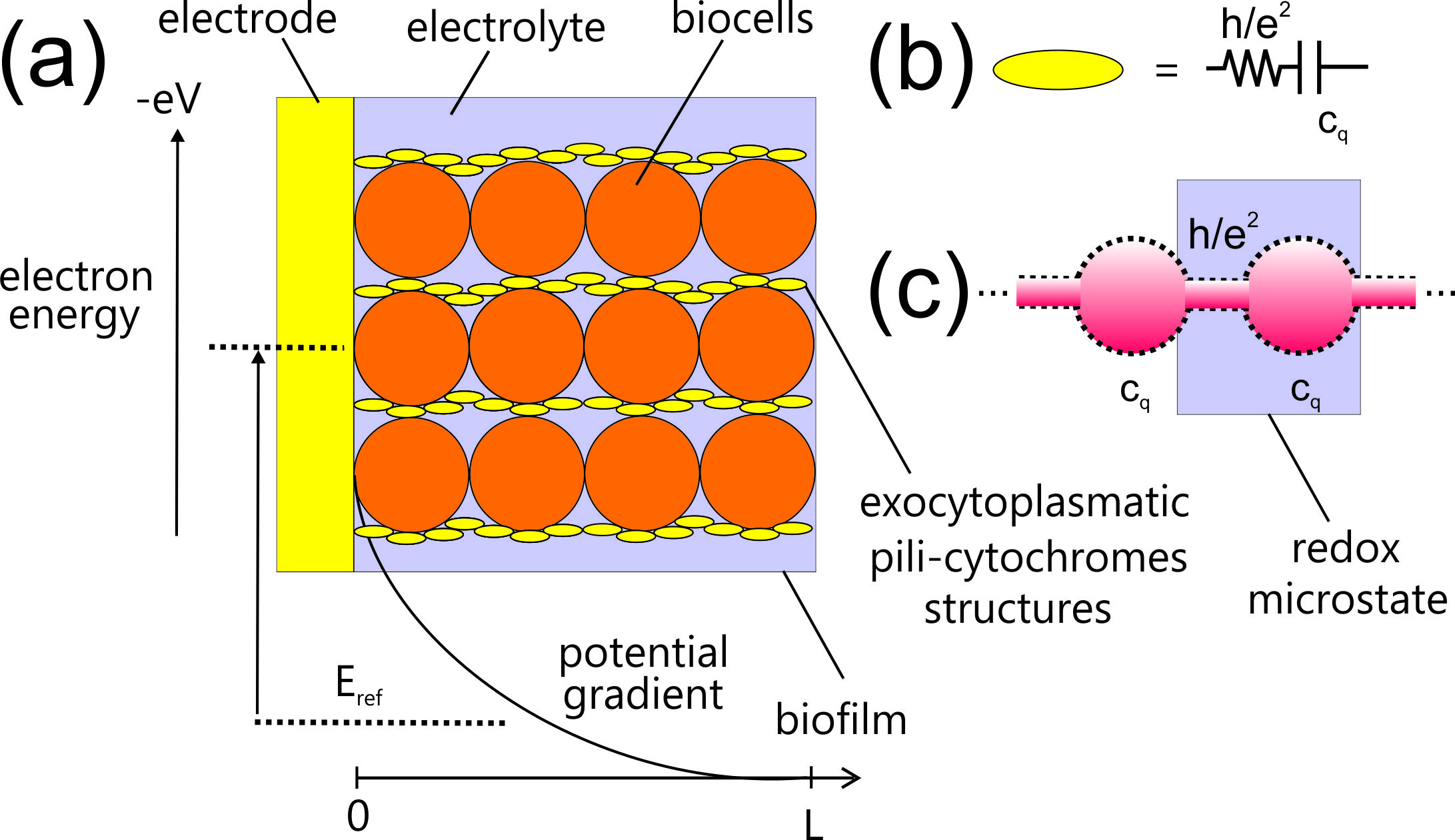}
\caption{(a) Schematic of a biofilm within cells interconnected with the electrode using a pili-cytochrome-based exocytoplasmatic structures as pictured in (b). In (b), the molecular characteristics of these structures form a microstate described by a mechanical statistical ensemble composed of a series combination of quantum resistors $h/e^2$ and capacitors $C_{q}$. In (c), the connection between these molecular statistical microstates is depicted. This electric structure allows electrons to be transported through the capacitors because of the quantum character of the series combination of conductive and capacitive elements of the ensemble. This transport is allowed through capacitors only because of its chemical character related to the HOMO and LUMO levels~\citep{Bueno-book-2018} associated with these types of capacitors(regarding more details of the role played by HOMO and LUMO states associated with the meaning of $C_q$, see SI document).}
\label{fig:biofilm}
\end{figure}

\begin{figure}[!t]
\centering
\includegraphics[scale=0.45] {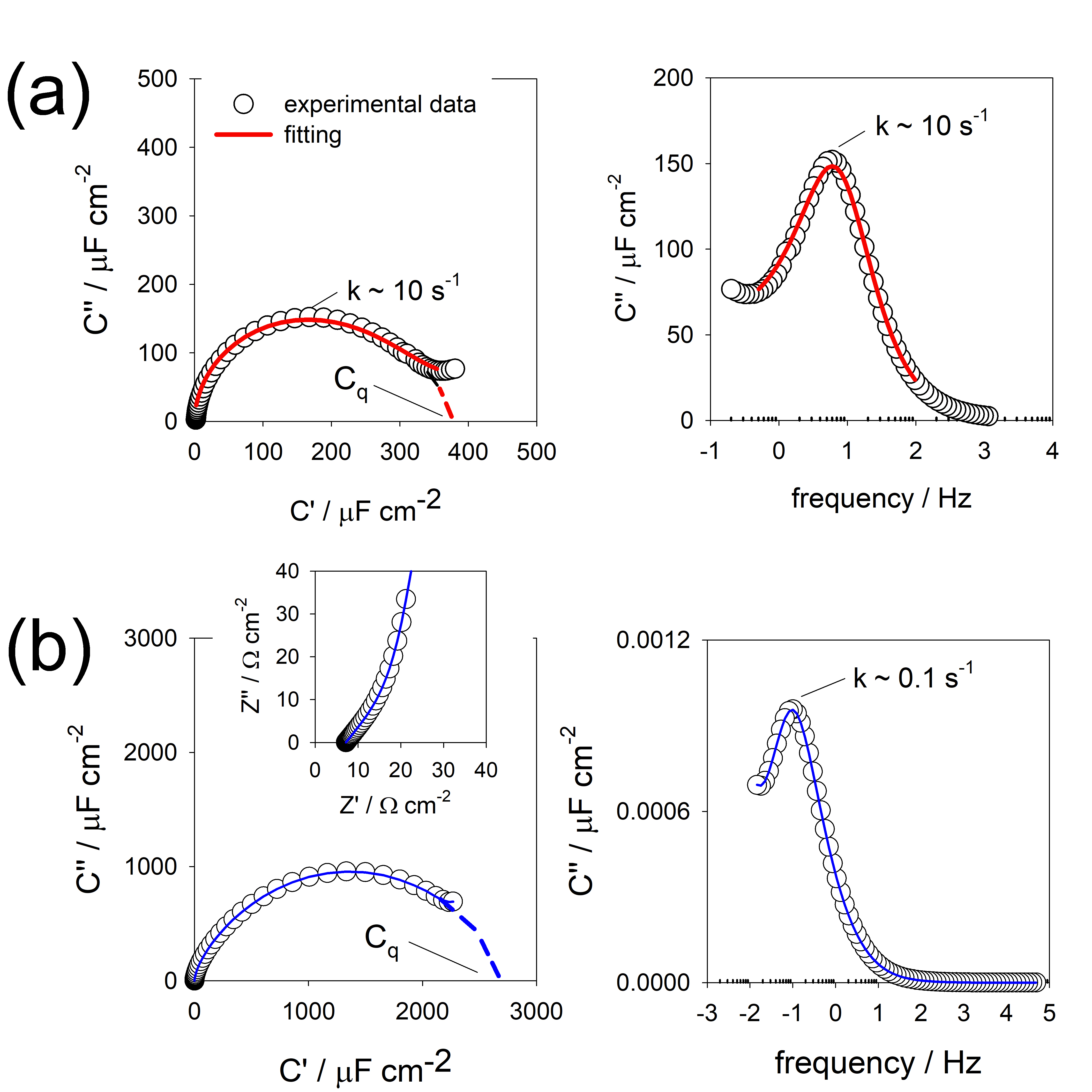}
\caption{(a) Capacitance spectra obtained from impedance spectroscopy for a (a) redox molecular film\cite{Bueno-2020, Bueno-book-2018,Bueno-2014} and a (b) biofilm of \textit{Geobacter sulfurreducens}\cite{Bueno-2015-Geobacter}. The equivalent circuit analysis and response are considerably similar in both films. Differences are associated with the fact that in (b), there is a potential decay in the response of the conductive-capacitive quantum rate states in the length $L$ of the biofilm. Further, in (b), the dependence of capacitance $C_q$ on the potential of the electrode follows a Boltzmann whereas in (a), it follows a Fermi–Dirac statistics. The inset in (b) shows the impedance response that identifies the potential decay, which is modeled as a transmission line (see Figure~\ref{fig:circuit-structure}\textit{b}). In both (a) and (b) situations, the electron transport complies with Laviron electrode kinetics\cite{Laviron-1979}, where the transport is diffusionless, in agreement with the quantum electrodynamics description the electron transport.}
\label{fig:ECS-spectra}
\end{figure}

Therefore, with (adiabatic) or without (non-adiabatic) a perfect transmittance mode of the quantum channels, the quantum electrodynamics is kept and the interpretation of the long-range electron transport within the quantum rate theory can be investigated in complex biological systems such as Geobacter films~\citep{Bueno-2015-Geobacter}, besides redox-active monolayers~\citep{Bueno-2020-ET, Alarcon-2021, Bueno-2023-3}, molecular junctions~\citep{Bueno-2023-ICT} and graphene~\citep{Bueno-2022}.

\begin{figure}[!t]
\centering
\includegraphics[scale=0.38] {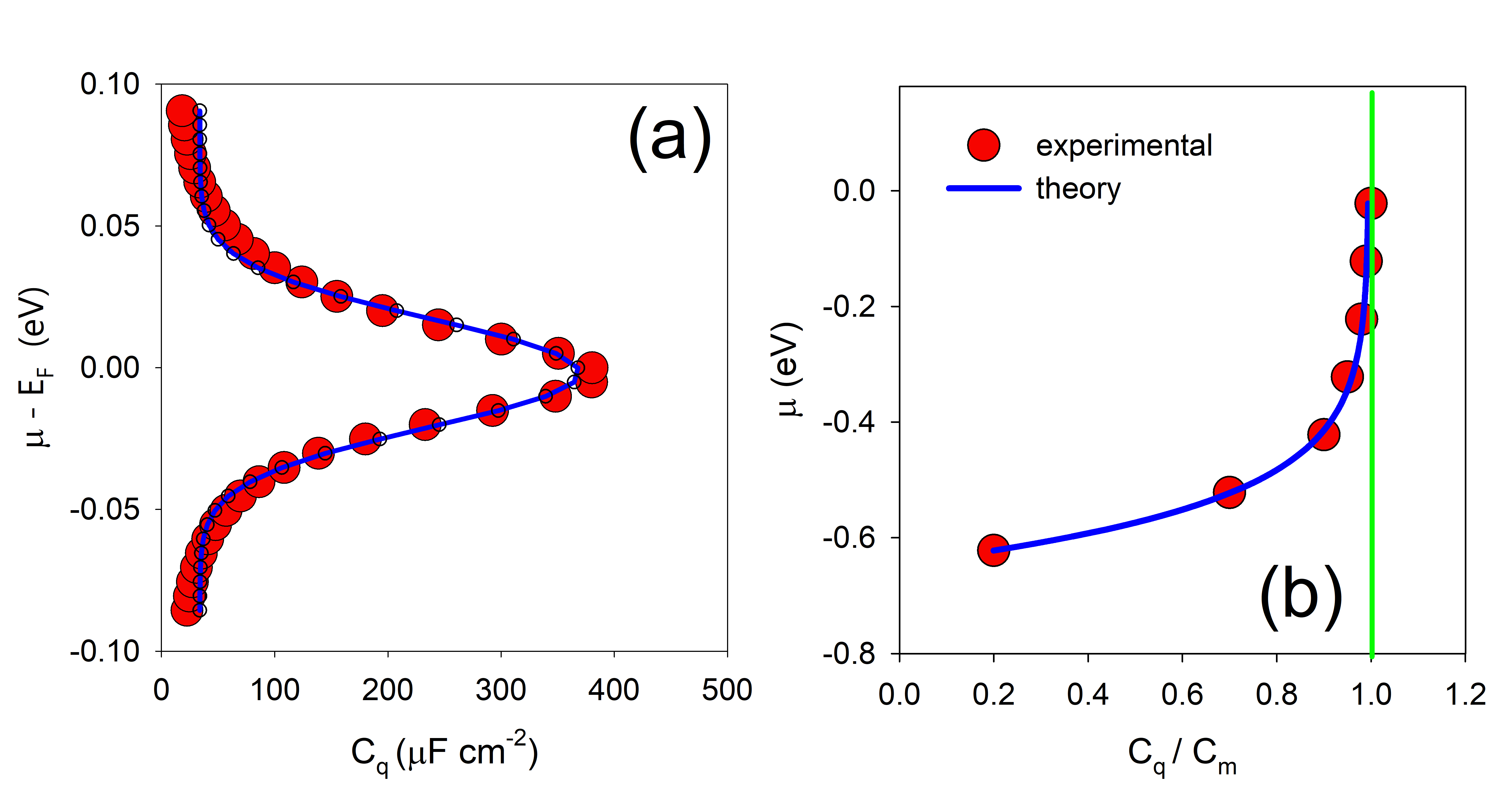}
\caption{Capacitance is proportional to the redox density-of-states that can be calculated in both configurations, i.e. in (a) a redox molecular film (depicted in Figure~\ref{fig:molecular-film} for an 11-ferrocenyl-undecanethiol monolayer self-assembled on a gold electrode) and in (b) a biofilm (depicted in Figure~\ref{fig:biofilm}). In both molecular film and biofilm scenarios, there is an equivalent quantum RC electrodynamics with different statistical mechanics, where the capacitance is modeled as $C_{q} = \left( e^2N/{k_{B}T} \right) f(1-f)$ in both cases. However, in (b), $f$ is approximated to the Boltzmann distribution, i.e. $f = \exp \left( -\Delta E /k_{B}T \right)$ to conform with the long-range pathway of length $L$ of the biofilm.}
\label{fig:DOS}
\end{figure}

The energy $E = e^2/C_q$~\cite{Bueno-2023-3, Bueno-2018, Bueno-book-2018} can be obtained in both molecular size (nanometre thick, see Figure~\ref{fig:molecular-film}\textit{a}) and biological (micrometer thick, see Figure~\ref{fig:biofilm}\textit{a}) films as a matter of comparison and strong experimental evidence of the applicability of the same theory in two different systems. The measurement of $C_{q,0}$, at lower frequencies (representing the charge equilibrium condition of the measurement), is equivalently achieved in both systems through complex capacitive diagrams, as shown in Figure~\ref{fig:ECS-spectra}, for molecular (Figure~\ref{fig:ECS-spectra}\textit{a}) and biological (Geobacter) films (Figure~\ref{fig:ECS-spectra}\textit{b}).

The measurement of $C_q$ in molecular films within a thickness (length of electron transport) $l$ of about 1.8 nm (11-ferroceny-undecanethiol) considering a non-adiabatic electron coupling term of $\kappa = \exp \left({-\beta l} \right) = G/NG_0$ has been demonstrated previously~\citep{Sanchez-2022-1}. The goal here is to illustrate the applicability of the quantum rate model to this thiolated molecular film (Figure~\ref{fig:molecular-film}\textit{a}) in comparison with Geobacter biofilm. The $\beta$ and $l$ terms of the electron coupling of the molecular film are obtained (or inferred) independently, as discussed in reference~\citep{Sanchez-2022-1}, from where it was demonstrated a good agreement ($\sim$ 3\% of differences) between the parameters inferred from a capacitive-derived impedance analysis and that obtained independently. For instance, $G/G_0$ and $N$ are obtained from impedance or complex capacitance spectra, as shown in Figure~\ref{fig:ECS-spectra}, where at the Fermi-level of the interface, $N = 4 k_B T C_q/e^2$ is directly accessible from the value of $C_q$.

The certainty of the quantum rate model is also evaluated from the measurement of $C_q$ as a function of the energy of the electrode, where it can be shown that $C_q$ follows a Gaussian-like shape (Figure~\ref{fig:DOS}\textit{a}), predicted by a thermally dependent capacitance of Eq.~\ref{eq:Cq-thermal}. The agreement between the DOS shape (obtained from $C_q$) measured for the molecular film to the theory is definitively evaluated in Figure~\ref{fig:DOS}\textit{a}, where the mathematical fitting of the experimental curve (red dots) to the model (blue line), as predicted by Eq.~\ref{eq:Cq-thermal}, is very good.

There are differences between molecular and biological film responses (see Figure~\ref{fig:DOS}), but the differences are not attributed to different quantum electrodynamics governing the electron transport in these films (that follows Eq.~\ref{eq:Complex-Cq}), but can be solely attributed to energy $e^2/C_q$ distribution and charge occupancy within the length of these films. Therefore, the differences are owing to statistical rather than quantum mechanics. In other words, as can be seen in Figure~\ref{fig:ECS-spectra}, where in Figure~\ref{fig:ECS-spectra}\textit{a} is the response of the molecular film whereas in Figure~\ref{fig:ECS-spectra}\textit{b} of the biological film, the quantum RC dynamics are equivalent. These quantum RC responses, each inferred from the measurement of these films at the Fermi level of the electrodes, show a similar quantum RC pattern that follows equivalent phenomenological quantum electrodynamics predicted by Eq.~\ref{eq:Complex-Cq}. Complex capacitive spectra of both films are different in magnitude but differ slightly in shape, and both are only differing regarding the magnitude of $C_q$ and of $k = 1/\tau$, but not in the phenomenological quantum RC response that fits quite well to Eq.~\ref{eq:Complex-Cq}.

The quantum RC characteristic of the molecular film is straightforwardly evaluated by taking the magnitude of $C_q$, which is $\sim$ 400 $\mu$F cm$^{-2}$. Considering an electro-active area of $\sim$ 0.039 cm$^2$, it leads to a capacitance of $\sim$ 15 $\mu$F, which according to Eq.~\ref{eq:G/C-degeneracy} corresponds to a $k$ of $g_e G_0/C_q$ $\sim$ 155 $\mu$S/15 $\mu$F $\sim$ 10 Hz, in agreement with the $k$ value obtained graphically in the Bode capacitive diagram of Figure~\ref{fig:ECS-spectra}\textit{a}, in the right-hand side. The fitting of the spectrum to Eq.~\ref{eq:Complex-Cq} is shown in the red line. This fitting confirms that $k = g_e/[2\pi R_q C_q] \sim$ 10 Hz, where $R_q$ encompasses the total series resistance of the interface, as better discussed in references~\citep{Sanchez-2022-1,Bueno-2023-3}. 

The magnitude of $C_q$ for the biological film is 2,800 $\mu$F cm$^{-2}$ with a characteristic RC frequency of 10 s, corresponding to a $k$ value of 0.1 Hz. Note that both 400 and 2,800 $\mu$F cm$^{-2}$ of capacitances correspond to values that are commonly referred to as pseudo-capacitances~\cite{Bueno-2019} that are hundreds of times higher than dielectric double-layer like non-faradaic capacitive responses, typically not higher than 20 $\mu$F cm$^{-2}$. The pseudo-capacitance response of these films is explained by the quantum rate model and cannot be associated with the Coulombic separation of charge phenomenon~\citep{Bueno-2019,Bueno-2023-3,Bueno+Davis-2020}.

\begin{figure}[!t]
\centering
\includegraphics[scale=0.65] {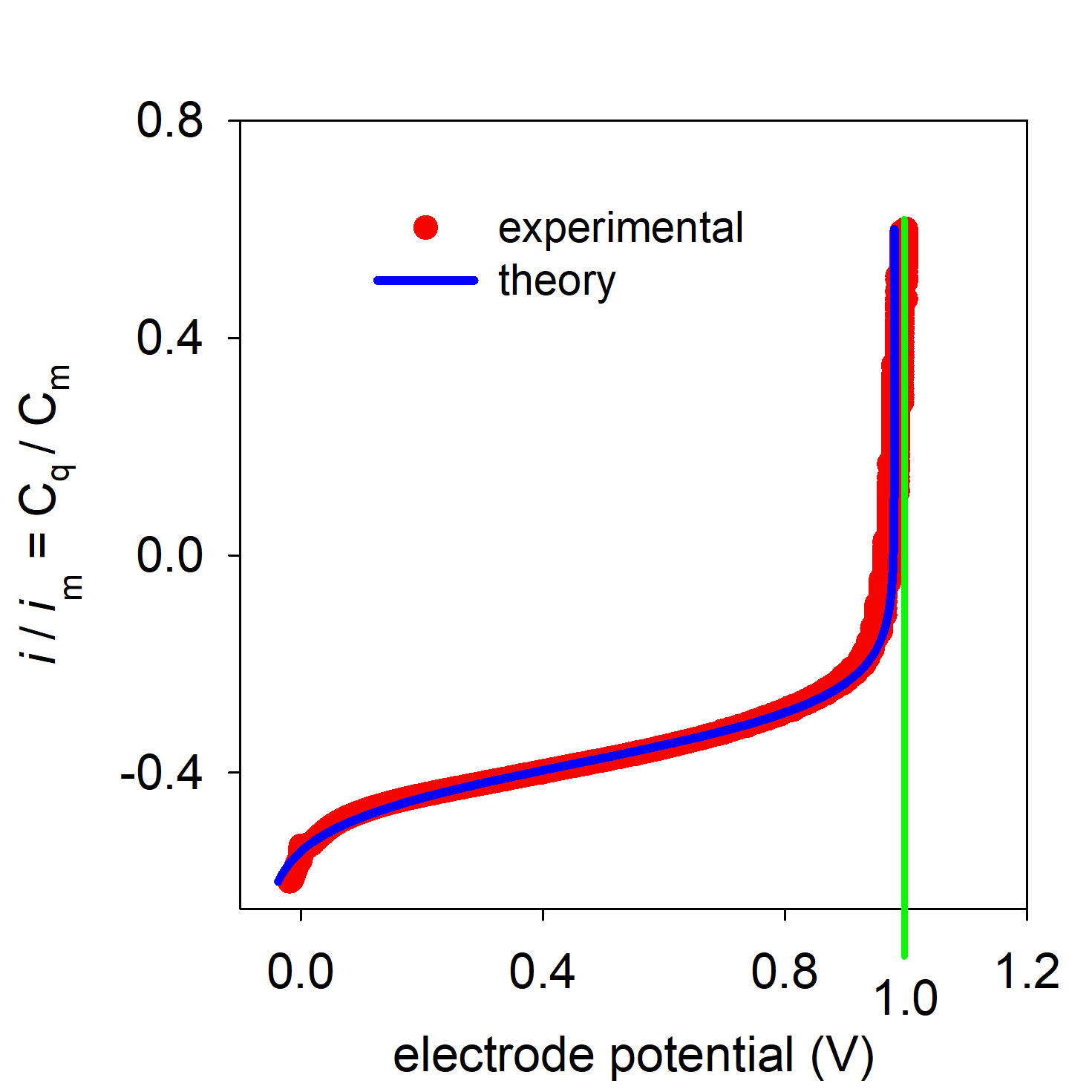}
\caption{Normalised electric current of a \textit{Geobacter sulfurreducens} biofilm scanned in equilibrium (lower scan rate) as a function of the electrode potential. $i_m$ denotes the maximum electric current achieved for the maximum chemical or quantum capacitance value $C_m$. This is the direct current test of the accuracy of the quantum rate model. The experimental curve is fitted to the theoretical model of the current $i = C{_q}s$, where $C_{q} = \left( e^2N/{k_{B}T} \right) \exp \left( -\Delta E /k_{B}T \right)$ and $s = V/\tau$. Because $\tau = R_{ct}C_{q}$, where $G = 1/R_{ct}$, $i = GV$ and the electric current can be alternatively $i = C{_q}s$ written as $i = GV$, i.e., as a function of $G = \kappa G_0 N$ or $C_{q} = \left( e^2N/{k_{B}T} \right) \exp \left( -\Delta E /k_{B}T \right)$.}
\label{fig:I-V}
\end{figure}

Accordingly, the electric current associated with charging these pseudo-capacitive states is not governed by classical laws of physics. The electric current which is $i_0$ at the Fermi level of the electrode, follows a time-dependent dynamics, in agreement with the quantum rate theory, predicting relativistic quantum dynamics. Generally, this implies that $i = C_q s$, where $s = dV/dt$ is the potential perturbation in time, which in a current-voltage ($i-V$) plot $s$ corresponds to the scan rate. In the quantum rate model interpretation, $s = V/\tau = V k$, which corresponds to $i = C_q (V/R_q C_q) = V/R_q$ or $G = i/V = 1/R_q$, in agreement to the analysis conducted in the previous paragraph and to the quantum RC dynamics interpretation of Figure~\ref{fig:ECS-spectra}.

Following the analysis of $i = C_q s$ and noting that quantum RC dynamics is obeyed in both molecular and biological films, the distribution of energy $E = e^2/C_q$ within a quite different thickness must be distinct. Hence, the electric current $i$ is predicted to follow a similar rate of electrodynamics owing to the equivalent quantum RC phenomenology, but with different charge occupancy dynamics governed by different thermal statistics in these films. 

Indeed, the DOS of the molecular film follows a shape that obeys, as depicted in Figure~\ref{fig:molecular-film}, a thermal distribution and a statistics of occupation $f$ that follows Eq.~\ref{eq:Cq-thermal}, where $f = n/N = \left(1 + \exp  \left( \Delta E /k_{B}T \right) \right)^{-1}$ obeys a Fermi-Dirac occupancy probability. For instance, in terms of redox reaction dynamics, this corresponds to a minimum (null) $n = f N$ occupancy in which $f = 0$ when the film is fully oxidized, whereas a maximum $n = N$ occupancy occurs when the film is fully reduced for which $f = 1$.

\begin{figure*}[!t]
\centering
\includegraphics[scale=0.60] {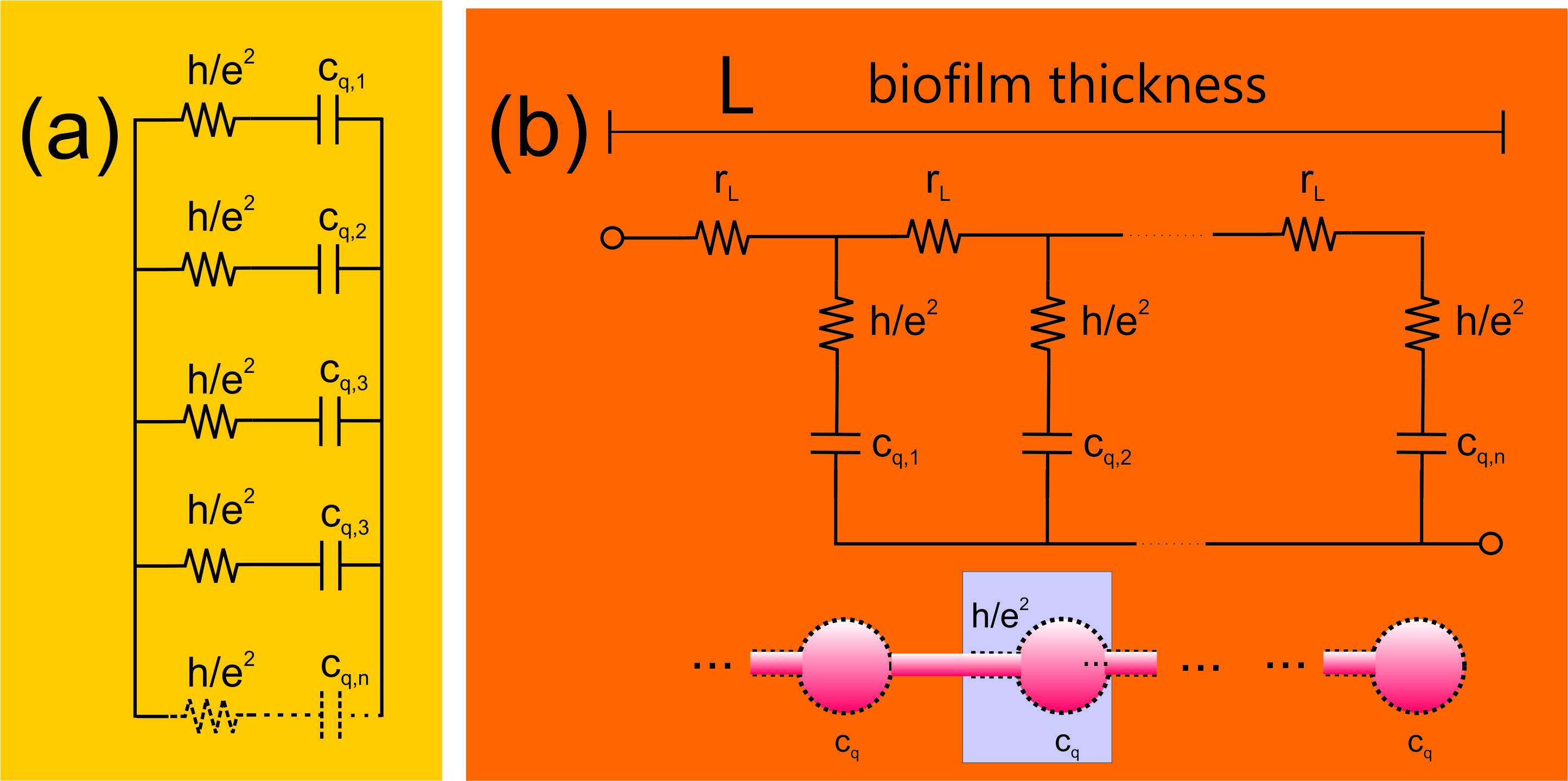}
\caption{(a) The equivalent circuit within resistive $h/e^2$ and capacitive $c_{q,n}$ microstate elements of a redox monolayer, as depicted in Figure~\ref{fig:molecular-film}. (b) Equivalent circuit for the biofilm. In (b), the resistive $h/e^2$ and capacitive $c_{q,n}$ individual elements of the circuit ensure that the electric current is distributed along the length $L$ following a transmission line scheme. The transmission line accounts for experimentally observed potential decay $L$ in agreement with the scheme depicted in Figure~\ref{fig:biofilm}\textit{a}. This is in agreement with the average value of the capacitance $C_{q} = \left( e^2N/{k_{B}T} \right) \exp \left( -\Delta E /k_{B}T \right)$, where the electrode potential energy is accounted as $\Delta E = -eV = \mu - E_F$, hence taking $E_F$ as a referential energy level.}
\label{fig:circuit-structure}
\end{figure*}

For both fully oxidised ($f = 0$) or reduced ($f = 1$) states of occupancy, $C_q = \left( e^2N/{k_{B}T} \right) f(1-f)$ minimises. At the Fermi level of the electrode, corresponding to $f = 1/2$, $C_q = e^2 N/4 k_{B}T$ maximizes, corresponding to half of the $n$ occupancy in which both the oxidized and reduced states of the electrode are equal and the electric ambipolar current maximizes to $i_0$, as better discussed in section~\ref{sec:QED-SMech}, and demonstrated in Figure~\ref{fig:DOS}\textit{a}.

Now it is possible to demonstrate that the differences between molecular and biological films correspond to a distinction between energy occupancy $E = e^2/C_q$ and electron thermal statistical distribution that is modulated by $C_q$ within its related statistical mechanical settings. In the molecular film, $f$ statistical occupancy follows a Fermi-Dirac whereas, in the biological film, it follows Boltzmann statistics (fitting of the experimental data to the model is shown in Figure~\ref{fig:DOS}\textit{b}). In other words, $f = n/N$ equates to $\left(1 + \exp  \left( \Delta E /k_{B}T \right) \right)^{-1}$ in the molecular film and to $\exp  \left( -\Delta E /k_{B}T \right)$ in the biological film, as shown in Figure~\ref{fig:DOS}. As Boltzmann is a particular setting of the Fermi-Dirac distribution, it is proved that the only differences between the electrodynamics within these films are owing to statistical mechanical settings rather than to the electrodynamics that follows quantum RC relativistic characteristics, as discussed in section~\ref{sec:QED-SMech} and demonstrated in the capacitive spectra of Figure~\ref{fig:ECS-spectra}.

Therefore, whether the Boltzmann approximation is applied to Eq.~\ref{eq:Cq-thermal}, i.e., under conditions where $\exp (\Delta E/k_{B}T) >> 1$ and $f \sim \exp  \left( -\Delta E/k_{B}T \right)$, the $C_q$ response, and consequently the $i = C_q s$, measured at the interface for the biological film, can be modeled through $\left( e^2N/k_{B}T \right) \exp \left( -\Delta E / k_{B}T \right)$, where the pre-exponential term corresponds to a maximum capacitance value of $C_m = e^2N/k_{B}T$, which is reached whenever the potential energy of the electrode equates to Fermi level $E_F = \mu_D - \mu_A$ of $D$ and $A$ components of the statistical ensemble with $f = 1$ and $n = N$ for the biofilm. The adjust (blue line) of $C_q \sim \left( e^2N/k_{B}T \right) \exp \left( -\Delta E / k_{B}T \right)$ to the experimental data (red dots) is considerably well, as shown in Figure~\ref{fig:DOS}\textit{b}. Note that the $C_q$ function used for modelling the molecular film, as shown Figure~\ref{fig:molecular-film}\textit{a}, encompasses the $C_q$ of the biofilm, as a particular thermal statistical distribution setting of Eq.~\ref{eq:Cq-thermal}.

Note also that $C_q = \left( e^2N/4{k_{B}T} \right) \exp \left( -\Delta E / k_{B}T \right)$ reaches a maximum $C_m = e^2N/{k_{B}T}$ value; a limit for the biofilm capacitance indicated by the green line in Figure~\ref{fig:I-V}, which is achieved whenever the potential energy of the electrode equates to the Fermi level $E_F$ of the biofilm~\citep{Bueno-2015-Geobacter}. This is a constant limiting value such as $C_q \sim C_m$ that is interpreted as the maximum charge occupancy of the biofilm~\cite{Bueno-2015-Geobacter}. This occurs whenever the redox sites of the biofilm are fully oxidized.

Following the analysis conducted by considering $C_q = \left( e^2N/4{k_{B}T} \right) \exp \left( -\Delta E / k_{B}T \right)$, the $i-V$ curve of the biofilm, as depicted in Figure~\ref{fig:I-V}, can be adjusted to the theory by considering the charge state of the biofilm as $q = C_qV$, where the derivative as a function of time provides $i = C_{q}s$, with $s = dV/dt$ as scan rate of electric potential, from which the $i-V$ curve is obtained. A very good fitting of the experimental data (red dots) to the theory (blue line) is demonstrated in Figure~\ref{fig:I-V}.

If the scan rate is sufficiently slow to ensure that the current is scanned in a charge equilibrium condition, then $s = V/\tau$, where $\tau = R_q C_{q}$ and $R_q \sim R_{ct}$, as equivalently noted in the case of molecular film electrodynamics~\citep{Sanchez-2022-1}. By noting that $G = 1/R_{ct}$, this analysis complies with $G = k C_q$ and where $k = 1/\tau$, thereby providing an additional demonstration of the certainty of the theory. Hence, $G$ is finally obtained as $\kappa (e^2/h) N = \kappa G_0  N$, which is equivalent to $G(\mu) = G_0 \sum_{n=1}^{N} T_n(\mu)$, as demonstrated in a previous works~\citep{Bueno-2020-ET, Sanchez-2022-1}, where $\sum_{n=1}^{N} T_n(\mu) = \kappa N$. In other words, the conductance follows quantum electrodynamics in \textit{Geobacter sulfurreducens} biological films in a similar way to that of molecular films~\citep{Bueno-2020-ET, Sanchez-2022-1} and the properties of both films, astonishingly, follows a quantum RC coherent electrodynamics at room temperature.

The above analysis is consistent with a previous study in which it was demonstrated that the quantum mechanical efficiency for the transport of electrons from the electrolyte to the electrode remains with a quantum mechanical efficiency if the interface of the electrode is chemically modified with redox capacitive states that serve as intermediate energy levels~\citep{Sanchez-2022-1}. The use of classical mechanics reasoning for the latter assumption is counterintuitive, and the transport of electrons within the ET reactions context is performed under a diffusionless condition, which is also the case of the biofilm, as demonstrated here. 

Accordingly, the respiration of Geobacter occurs with a diffusionless setting that resembles that of molecularly modified metallic electrodes that have proven to be better for probing redox reactions than the direct contact of a metallic probe with redox probe free in the solution phase, permitting a better quantum mechanical dynamics for the flowing of electrons from a solution phase redox states towards the electrode and \textit{vice-versa}, which is similarly observed in the transport of the electrons in the biological films, as studied here. Consistently, the presence of capacitive (see inset in Figure~\ref{fig:circuit-structure}\textit{b}) interconnected states intermediating the electron transport within the biological film permits electrons to be transmitted with a charge transfer resistance $\propto e^2/h$, leading to a maximum electrode-mediated quantum rate efficiency. This phenomenon can only be understood from a relativistic quantum mechanical perspective of electron transport, as discussed in section~\ref{sec:QED+ETr}.

To consider the long-range transport of electrons within the size of the biological film where there is a potential decay within the length $L$, a transmission line of quantized RC terms was considered, as shown in Figure~\ref{fig:circuit-structure}\textit{b}, as a precise way of considering the impedance response of the biofilm at higher frequencies (see inset of Figure~\ref{fig:ECS-spectra}\textit{a}). Hence, in the biological film, the RC ensemble of circuit elements is distributed within the thickness $L$ of the film and can be biologically associated with or interpreted as cytochrome redox states that are connected through nanowires with $\kappa e^2/h$ `metallic-like'\footnote{Note that, according to the quantum rate theory, the only difference between coherent (ballistic) and non-coherent (hopping) rate or conductance of which electrons are transmitted in the electrolyte embedded medium is settled by the meaning of $\sum_{n=1}^{N} T_n(\mu) = \kappa N$. The meaning of this analysis of the quantum rate theory is that the rate can be modelled differently according to $\nu = \kappa g_eG_0/C_q$, where a coherent or incoherent electron transport (or rate) is only a matter of how $\kappa$ is defined in a statistical scattering terminology within $\sum_{n=1}^{N} T_n(\mu)$. In other words, the adiabatic or non-adiabatic scattering in different degrees of the quantum rate does not determine the nature of the quantum electrodynamics that defines the electron rate or transport phenomenon, which remains the same independently of the different modes of transport operating in the distinct situations that can arise in biological systems.} conductances. The estimated value of the nanometre wiring length $L$ of each interconnect RC element of the transmission line was between 7 to 9 nm which can be interpreted as the action of the pili nanofilaments or an equivalent nanowire path environment\citep{Dahl-2022}. This long-range length wiring of $D$ and $A$ states through a molecular bridge such as pili nanofilaments or polymerized cytochrome OmcS path\citep{Dahl-2022} was also demonstrated to be possible using redox-modified DNA bridges~\citep{Ribeiro-2016}, where DNA strands perform as nanowires. Indeed, it was demonstrated that different DNA wire strands possess different electron transmittances~\citep{Ribeiro-2016}. 

The limiting oxidation state of the biofilm, as shown by the green line in Figure~\ref{fig:I-V}, corresponds to the extraction of electrons from capacitive sites interconnected by $\propto e^2/h$ conductive channels. These interconnected capacitive levels form a one-dimensional energy level path for electrons within $\propto g_s e^2/h$ conductivities depending on an assumed $\kappa$ adiabatic degree or an appropriate electron coupling for electron hopping mode of transport, thus forming a structure similar to that of a conducting band of semiconductors. This electronic structure interpretation of long-range electron transport is based on a particular setting of Eq.~\ref{eq:G/C} or \ref{eq:k-finiteT} for the biofilm and hence it agrees with the quantum rate theory analysis. This analysis effectively unifies controversial electronics and electrochemistry viewpoints on the long-range transport of electrons in biological chains.

Within the quantum rate depiction of the long-range transport of electrons, it is particularly interesting noting that the transport of electrons (in an adiabatic or non-adiabatic regime) occurs at frequencies between 15 Hz down to 0.1 Hz (see Figure~\ref{fig:ECS-spectra}), which corresponds to the extremely low frequency of the electromagnetic spectrum, that is, a region in which the wavelength is particularly high. Frequencies in this region permit a multitude of possibilities for $\nu \sim c_*/\lambda = i_0/e$ in agreement with Eq.~\ref{eq:G-v_F} and~\ref{eq:i0}. These possibilities are due to the high degree of freedom that can be achieved for the quantum electrodynamics due to the electrolyte environment that permits $C_e \sim C_q$.

In other words, the electric-field screening governed by electrolyte dynamics allows the transport of electrons at frequencies that guarantee the maximum efficiency of electron transmittance (corresponding to the quantum limit of electron transport) in different electrolyte settings (ion and counter ions distances) and solvent environments (solvating conditions). The adjustment of the frequency of the electron transport for the maximum quantum efficiency is intrinsic to the high efficiency with which electricity is conducted in biological systems, a condition that cannot be easily attained by man-made solid-state electronics. Furthermore, this frequency phenomenon is close in magnitude to that in which electricity is transmitted by transmission lines power stations using the AC mode of transport which is required to decrease the loss of energy during the transmission of electrical energy at long distances. However, there is a crucial difference, in biological structures within intrinsic electrochemical electrolyte medium, the transmittance of energy follows quantum relativistic electrodynamics. 

Accordingly, the velocity $c_*$ of transmittance of electrons in Eq.~\ref{eq:G-v_F}, according to relativistic quantum mechanics, cannot be estimated from a fixed $L$ owing to $L \sim \lambda$ (for a single adiabatic state mode of electron transmittance), but it can be varied and easily adapted to different circumstances of the biological environments and electron coupling possibilities. This $c_*$ velocity does not follow a `metallic-like' characteristic and hence the use of this term is unfortunate to describe the quantum conductance of electrons within the Landauer mode of conductive transport or within a coherent or incoherent electron hopping mechanism that follows a massless Fermionic mode of transport, as described in section~\ref{sec:QED+ETr}. The latter is because the velocity of the transmittance of electrons (following a coherent or incoherent mode of transmittance) in molecular structures is far similar to be likewise that of metals which are governed by a drift velocity with loss of energy that is quite inefficient compared to that observed in the mode of electron transport at the Fermi level.

In summary, the quantum conductance with the quantum rate mode permits electrons to `hop' from one $D$ to another $A$ state with a maximum relativistic quantum mechanical efficiency that depends on the $\nu = e^2/hC_q$ rate within a coherent or incoherent electron coupling bridge characteristic that complies with different situations without a loss of quantum efficiency, always obeying the phenomenology of Eq.~\ref{eq:Planck-Einstein}, even for different statistical mechanical settings imposed by a Fermi-Dirac or Boltzmann energy distribution of the DOS associated to $C_q$. The bridge, within a length $L$ in an ETrC scheme of electron transport, permits electrons to be transported in different chemical environments comprising long distances if required, as is the case demonstrated here, which plausibly (and without inconsistency between chemical and physical concepts) explain the respiration chain of Geobacter films.

\section{Final Remarks and Conclusions}

It was demonstrated that $C_q$, which determines the energy properties $E = e^2/C_q$ of different molecular or biological films, governs the electron occupancy of the states in both molecular or biological films. Nonetheless, the electron occupancy in each case follows distinct thermal and electron scattering statistics albeit within an equivalent, in nature, quantum RC electrodynamics. In both scenarios, the electrodynamics follows quantum mechanical rules with different energy occupancy that depends only on the statistical mechanics settings ruled by $C_{q}$, attending the different length settings of molecular and biological films. Ultimately, electrons follow a Fermi-Dirac occupancy of states $n = N\left(1 + \exp  \left( \Delta E/k_{B}T \right) \right)^{-1}$ in molecular films with a thickness length no longer than 2 nm and a Boltzmann occupancy of $n = N \exp  \left( -\Delta E /k_{B}T \right)$ in biofilms within a thickness of about 50 $\mu$m.

In conclusion, it was demonstrated that the long-range electron transport in an electrochemical medium that governs the respiration mechanism, as illustrated in the case of \textit{Geobacter sulfurreducens} biofilms physiology, are intrinsically controllable by equivalent quantum electrodynamics to that observed for short-range electron transfer processes within solely different statistical mechanics settings. Additionally to consider different thermal statistics settings, the quantum rate model is also able to take into account different coherent and incoherent modes or `metallic-like' transport \textit{versus} super-exchange mechanisms of electron transmission depending only on electron scattering statistics, which can be regarded to the transmission matrix $\sum_{n=1}^{N} T_n(\mu)$ component of the rate equation (Eq.~\ref{eq:k-finiteT}). Independent of different thermal or electron scattering statistics, the relativistic quantum electrodynamics setting of the rate remains operational.

This is a ground-breaking physical chemistry analysis that demonstrates how the efficiency of respiration -- a biological process responsible for energy conservation in living organisms -- is intrinsically quantum mechanical and modelable depending only on appropriate statistical thermodynamics (through the meaning of $C_q$) or electron scattering (through the meaning of $G$) requirements. Finally, it can be inferred that biological elements such as pili (\textit{per se} with a conductive ballistic character or not) and cytochrome (with quantum capacitive states) can behave differently or take distinct magnitudes according to distinct molecular setting structures in a variety of biological settings\footnote{It can exist within different molecular structures, but with similar electrodynamics, as described by Eq.~\ref{eq:G/C} or Eq.~\ref{eq:k-finiteT}.}; however, the meaning of the rate described by Eq.~\ref{eq:G/C} (in the zero-temperature limit) or Eq.~\ref{eq:k-finiteT} (in a finite-temperature description) likely would remain universal for the phenomenon of respiration and biological processes that requires long-range transport of electrons.

\section*{Conflict of Interest}
There are no conflicts of interest to declare.

\section*{Acknowledgements}

The author is grateful to Prof. Juan Pablo Busalmen for his helpful discussions and to the National Council for Scientific and Technological Development (CNPq) and Sao Paulo State Research Foundation (FAPESP, Grant no. 2017/24839-0) for the financial support of his research.

\bibliography{rsc}
\bibliographystyle{rsc}

\end{document}